\documentclass[%
 aip,
 amsmath,amssymb,
preprint,%
]{revtex4-1}

\usepackage{graphicx}
\usepackage{dcolumn}
\usepackage{bm}

\usepackage[utf8]{inputenc}
\usepackage[T1]{fontenc}
\usepackage{mathptmx}
\usepackage{etoolbox}

\makeatletter
\def\@email#1#2{%
 \endgroup
 \patchcmd{\titleblock@produce}
  {\frontmatter@RRAPformat}
  {\frontmatter@RRAPformat{\produce@RRAP{*#1\href{mailto:#2}{#2}}}\frontmatter@RRAPformat}
  {}{}
}%
\makeatother

\usepackage{caption}
\usepackage{subcaption}

\usepackage[free-standing-units=true]{siunitx}

\usepackage{xcolor}

\definecolor{ao}{rgb}{0.0, 0.5, 0.0}

\usepackage{xspace} 
\newcommand{\etal}{\textit{et~al.}} 
\newcommand{\etalnp}{\textit{et~al}} 
\newcommand{\via}{\textit{via}\xspace} 

\begin{document}

\preprint{AIP/123-QED}

\title{Numerical and Experimental Study on the Addition of Surface Roughness to Micro-Propellers}
\author{J.P. Cooke}
\author{M.F. Campbell}%
\author{E.B. Steager}
\author{I. Bargatin}
\author{M.H. Yim}
\author{G.I. Park}
 \email{gipark@seas.upenn.edu.}
\affiliation{ 
Department of Mechanical Engineering and Applied Mechanics, University of Pennsylvania, Philadelphia, PA, 19103
}%

\date{\today}

\begin{abstract}
Micro aerial vehicles are making a large impact in applications such as search-and-rescue, package delivery, and recreation. Unfortunately, these diminutive drones are  currently constrained to carrying small payloads, in large part because they use propellers optimized for larger aircraft and inviscid flow regimes. Fully realizing the potential of emerging microflyers requires next-generation propellers that are specifically designed for low-Reynolds number conditions and that include new features advantageous in highly viscous flows. One aspect that has received limited attention in the literature is the addition of roughness to propeller blades as a method of reducing drag and increasing thrust. To investigate this possibility, we used large eddy simulation to conduct a numerical investigation of smooth and rough propellers. Our results indicate that roughness produces a 2\% increase in thrust and a 5\% decrease in power relative to a baseline smooth propeller operating at the same Reynolds number of $Re_c=6500$, held constant by rotational speed. We corroborated our numerical findings using thrust-stand-based experiments of 3D-printed propellers identical to those of the numerical simulations. Our study confirms that surface roughness is an additional parameter within the design space for micro-propellers that will lead to unprecedented drone efficiencies and payloads. 
\end{abstract}

\maketitle

\section{Introduction}
\label{sec:intro}

Small unmanned aerial vehicles (UAVs) have long held important roles in situations that are unsafe for humans, including detection of bio-chemical hazards, aerial inspection and mapping of natural disaster zones, and data-collection on other planets~\cite{mueller2003aerodynamics,Smedresman2011Design,kroo2001mesicopter}. Thanks to advances in the technology, UAVs are shrinking to smaller length-scales, offering advantages in their simplicity, multiplicity, and ability to access small spaces~\cite{mueller2003aerodynamics, Smedresman2011Design}. This subset of vehicles are known as micro aerial vehicles (MAVs), and, as defined by the Defense Advanced Research Projects Agency (DARPA), cannot exceed 6~in (15~\si{\centi\meter}) in any dimension~\cite{Deters2018-4122}. A critical challenge facing this new class of drones is coping with low Reynolds number ($Re$) aerodynamics, where viscous forces play a significant role.  As of the writing of this article, much of the existing body of aerodynamic knowledge related to MAVs concerns higher Reynolds numbers~\cite{mueller2003aerodynamics,Smedresman2011Design,kroo2001mesicopter}. The low-$Re$ flow regime, which is defined by $Re = O(10^5)$ and below, is known to be highly viscous in nature, resulting in thicker boundary layers which increase drag and decrease lift~\cite{kroo2001mesicopter}. Accordingly, MAV propeller design and manufacturing should account for viscous effects to better overcome losses in lift. There have been some works investigating commercially available small-scale propellers, ranging in diameter from approximately 6-12~\si{\centi\meter}~\cite{deters2008static, deters2014reynolds, selig2011wind}, and a recent study by Deters~\etal~\cite{Deters2018-4122} looked at propellers in the range of 3-7~\si{\centi\meter} operating at $Re_c$ numbers 10,000 and greater. These studies are useful for understanding the current performance level of available low-$Re$ propellers.

Previous researchers have looked to bio-inspired, rough surfaces to enhance airfoils and propellers at lower Reynolds numbers. Sharkskin-like roughness was used on airfoils at $Re \sim O(10^5)$ to reduce drag~\cite{devey2020experimental}, and leading-edge tubercles also demonstrated efficiency improvements for propellers operating at $Re \sim O(10^4)$ and above~\cite{asghar2020application,butt2019numerical}. There have also been investigations on the addition of more simple roughness geometries on airfoils and wings operating at $Re \sim O(10^4-10^5)$~\cite{zhou2012effects,lee2005control,mukherjee2021corrugation}. Lee and Jang~\cite{lee2005control} found evidence of drag-reduction for airfoils by applying micro-riblet films, with better drag reduction as Reynolds number decreased. Further, roughness elements deployed as vortex generators  applied to low-$Re$ airfoils have showcased an ability to enhance lift at higher angles of attack, by altering pressure distribution, delaying separation, and through augmentation of the leading-edge vortex (LEV)~\cite{heine2013dynamic,tavernier2021controlling}. It has been shown for low-$Re$ flows that the LEV can enhance lift capabilities~\cite{lentink2009leading, ford2013lift, lee2017flight, jardin2021empirical}, and rotating wings are especially at an advantage due to stability from Coriolis forces~\cite{lee2017flight,jardin2017Coriolis}. We believe there is great potential for surface roughness to improve MAV propeller performance at lower length-scales and velocity-scales. Importantly, although some prior works have examined airfoil roughness in the upper end of the low-$Re$ regime, these illustrate the effects of roughness on flows involving translational, rather than rotational, motion. To our knowledge, our work is the first published study on textured MAV propellers. 

In this article we present a numerical and experimental investigation, well within the low-$Re$ regime, to explore the addition of surface roughness to a MAV propeller. We began with a baseline 4~\si{\centi\meter} diameter propeller, altered it by adding a simple surface texture pattern on the upper surfaces of its blades, and studied both the smooth and textured versions numerically at $Re_c=6500$ and experimentally at $Re_c \sim O(10^3-10^4)$. Our numerical results will show that surface roughness increases thrust coefficient values, strengthens the LEV, and alters the blade surface pressure distribution. Moreover, our experimental tests corroborated our numerical findings that surface roughness increases the propeller's thrust coefficient and reduces its power coefficient.

\section{Propeller Performance and Design}
\label{sec:prop_performance_design}

\subsection{Performance Metrics}
\label{ssec:perfMetrics}

Propellers can be characterized by the $T$~[\si{\newton}] they produce and by the amount of $P$~[\si{\watt}] required to drive them~\cite{Deters2018-4122}. The value of thrust may be measured with an experimental test stand, as described in Section~\ref{ssec:expt}, or taken from forces measured in a computational calculation, as outlined in the numerical study in Section~\ref{sec:numMethod}. Similarly, values of $Q$~[\si{\newton\meter}] may be measured with a test stand, and moments may be measured in a numerical calculation in order to find $P$, given as 
\begin{equation}
    P = 2\pi{n}Q.
    \label{power}
\end{equation} 
Here, $n$~[\si{rev\per\second}] is the propeller's rotation speed.  Non-dimensional forms of $P$ and $T$ are ideal for comparing the performance of different propeller designs. Specifically, the power and thrust coefficients $C_P$ and $C_T$ are calculated \via Eqs.~\eqref{c_p} and~\eqref{c_t}, respectively. 
\begin{equation}
    C_P = \frac{P}{\rho n^3 d^5}, 
    \label{c_p}
\end{equation}
\begin{equation}
    C_T = \frac{T}{\rho n^2 d^4}.
    \label{c_t}
\end{equation}
In this equation, $d$~[\si{\meter}] is the propeller diameter and $\rho$~[\si{\kilo\gram\per\meter\cubed}] is the air density. Pressure $p$~[\si{\pascal}] distributions across the blade surface provide insight into the presence and strength of vortex structures, such as the LEV. Further, these measurements may indicate potential separation and re-attachment points on the blade surface. A non-dimensional pressure coefficient $c_p$ (note the lowercase script and subscript) is useful for comparing the two propellers, and is given in Eqn. \eqref{c_pr}.  
\begin{equation}
    c_p = \frac{p-p_\infty}{\rho U^2_c}
    \label{c_pr}
\end{equation}
Here, $U_c$~[\si{\meter\per\second}] is the air velocity in the chord-wise direction at the corresponding blade-station, and $p_\infty$~[\si{\pascal}] is the far-field pressure. For the purposes of the experiments and numerical study, the chord-based $Re$ number is defined by 
\begin{equation}
    Re_c = \frac{\rho U_c c}{\mu},
    \label{Re_c}
\end{equation}
where the chord $c$~[\si{\meter}] at the 75\% blade-station is taken for the characteristic length scale, and the characteristic velocity is based on the rotational speed at the same position.  In the above, $\mu$~[\si{\newton\second\per\meter\squared}] is the air's dynamic viscosity. 

\subsection{Propeller Selection and Design}
\label{ssec:propDesign}

\begin{figure}[hbt!]
    \centering
    \begin{subfigure}[b]{0.75\textwidth}
        \centering
        \includegraphics[width=0.85\textwidth]{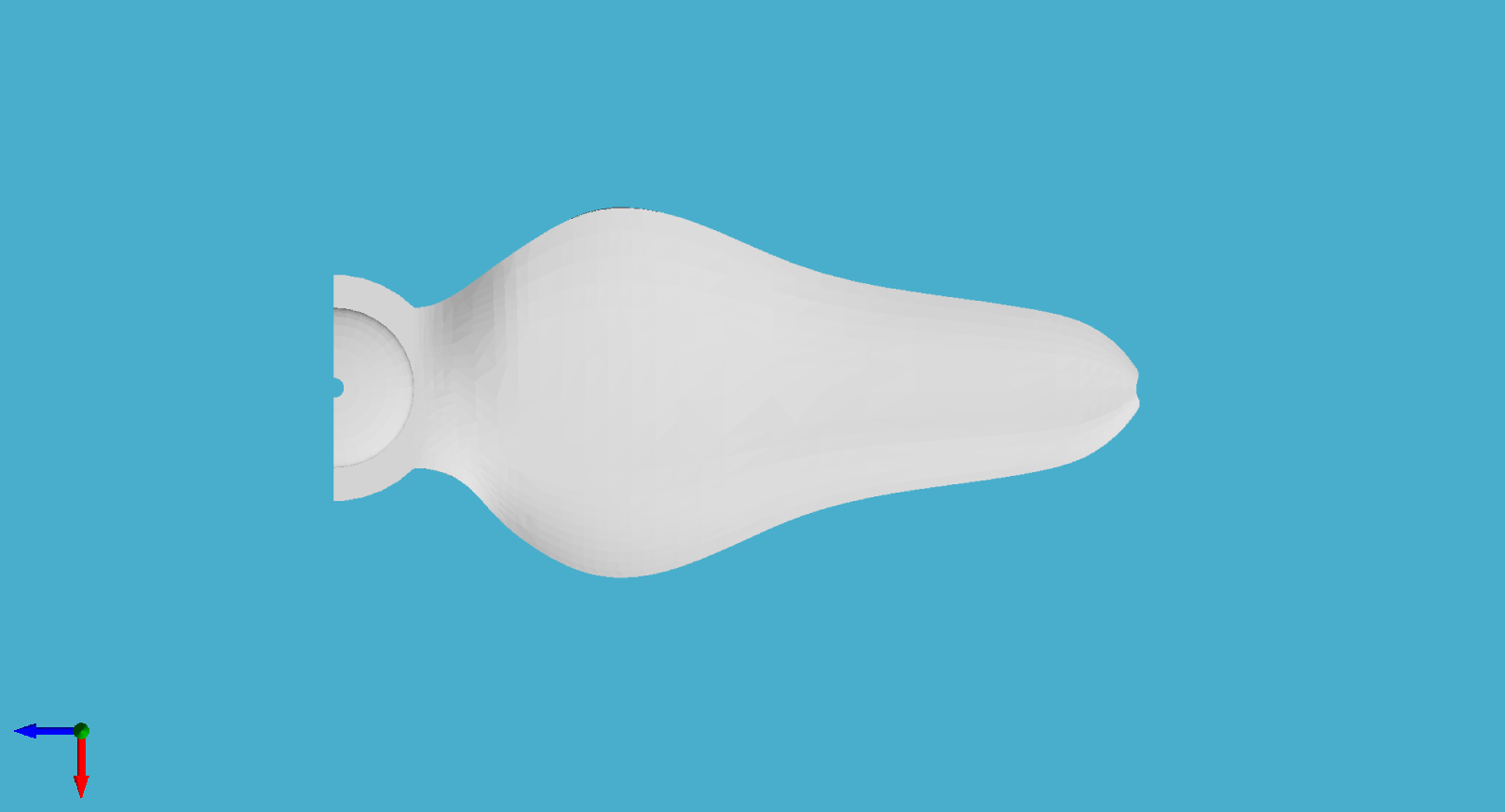}
        \caption{Smooth surface, baseline propeller.}
        \label{fig:smoothProp}
    \end{subfigure}
    \begin{subfigure}[b]{0.75\textwidth}
        \centering
        \includegraphics[width=0.85\textwidth]{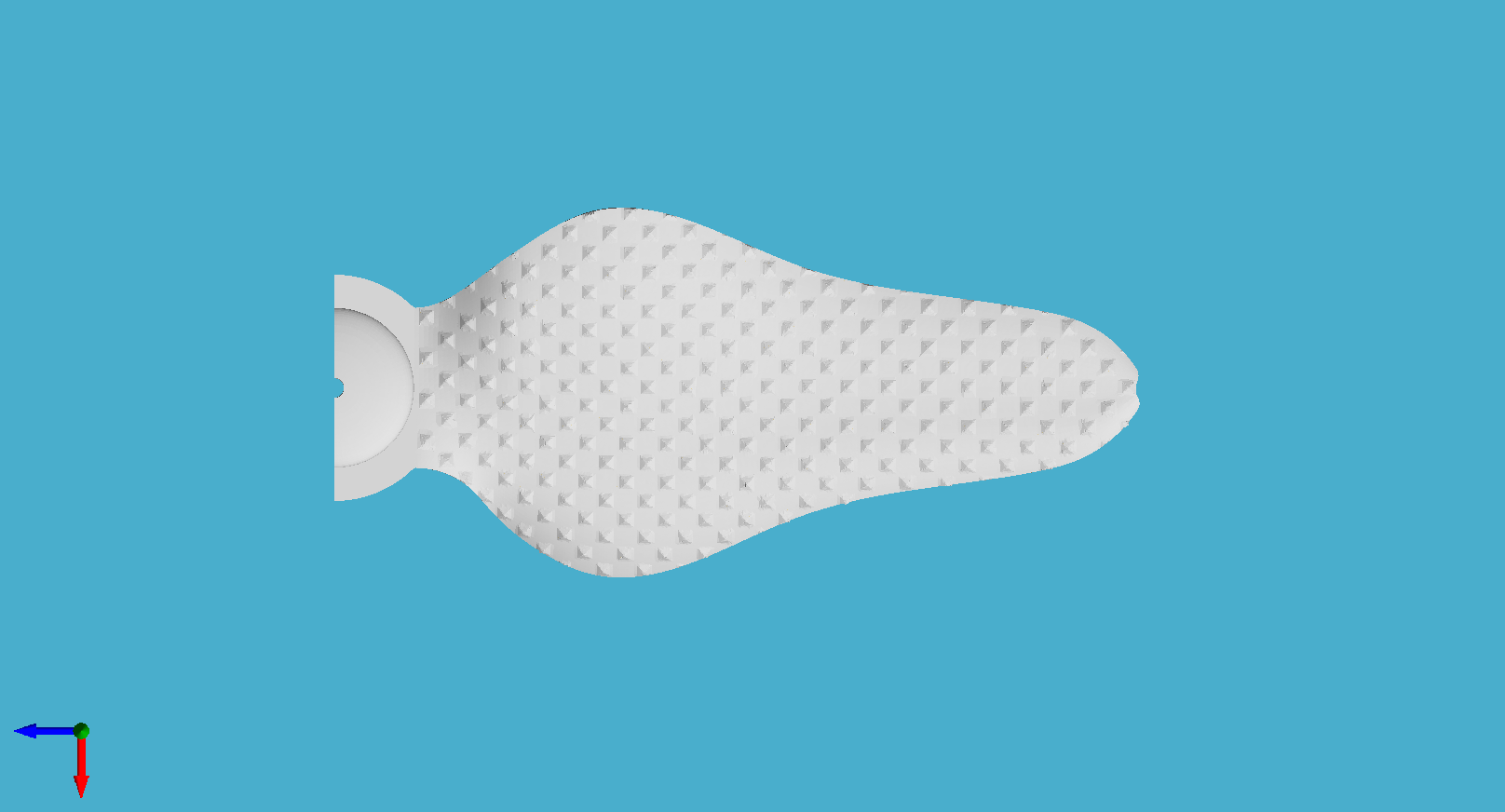}
        \caption{Propeller with surface roughness, shown by the textured pattern on the blade. }
        \label{fig:roughProp}
    \end{subfigure}
    \caption{CAD geometries of the smooth and rough surface propellers deployed in this investigation, note only half the symmetric propeller is shown.}
    \label{fig:props}
\end{figure}

For this investigation, we selected a propeller design similar to the one previously optimized by Kroo~\etal~\cite{kroo2001mesicopter} for MAVs at $Re_c = 2000$ (Figure~\ref{fig:smoothProp}). The 
report by Kroo~\etalnp.\  included angle of attack and chord length information as a function of radial position, as well as an airfoil profile. We simplified this design by employing a constant propeller thickness of 0.3~\si{\milli\meter} for the entire airfoil, taken by expanding the thickness $\pm0.15$~\si{\milli\meter} from the airfoil profile centerline. We implemented this modification and selected the $0.3$~\si{\milli\meter} thickness in particular to simplify the 3D printing of the propellers for the experiments. The diameter of our propeller is $d=4$~\si{\centi\meter}. It has two opposing blades with right-handed orientation. The propeller has a thick hub to help it grip the motor shaft for the experiments; we created it in SOLIDWORKS~2020. We note that other studies have also examined optimal airfoil profiles at low $Re$ (see Smedresman~\etal~\cite{Smedresman2011Design} and Ukken and Sivapragasmn~\cite{Ukken2019-130}), but these lack angle of attack and chord length information. We also note that Kroo~\etalnp.\ includes a design for a $Re_c = 6000$ propeller, where its optimal flow conditions more closely match those of this study. Our explicit goal in this work is not to create an optimal roughened propeller, but rather to explicate the impact of surface roughness on propeller performance. Selecting a propeller that has a room for improvement (by operating it outside of its optimal $Re_c$ range) was deemed appropriate in this sense.

We created a rough version of the baseline, smooth propeller by coating the entire topsides of the blades with a checkerboard array of 0.39~\si{\milli\meter}-side-rounded square bumps with $k=0.1~\si{\milli\meter}$ 
roughness height 
and a grid spacing of approximately 0.69~\si{\milli\meter} (see Figure~\ref{fig:roughProp}). {When comparing $k$ to the boundary layer thickness ($\delta$), the ratio $\delta/k$ is approximately 6.5. According to the criteria for k-type roughness of $\delta/k > 40$ 
\cite{jimenez2004turbulent}, 
this would characterize the roughness as being d-type, which protrudes further into the flow compared to the usual k-type roughness.} The features of the rough propeller are otherwise unchanged; the only difference between the smooth and rough propellers is the addition of this bump array. Our choice of this particular pattern and bump size is somewhat arbitrary because our goal in this work is not absolute propeller optimization. However, these bumps in particular offer several advantages: (1) the bumps are rounded and relatively short in height so that the roughened propeller still closely mimics the smooth propeller, meaning the propellers' flow patterns can be meaningfully compared; (2) the bumps are large enough to be easily discernible on the surface of the propeller such that flow features can be easily related to individual bumps; and (3) the bumps are large enough that they can be reliably 3D printed. 

\section{Numerical Method}
\label{sec:numMethod}

\subsection{Description of Numerical Study}
\label{ssec:descriptNS}

We analyzed both propellers numerically using the Navier-Stokes Solution to compare their performance with and without surface roughness. As mentioned above, both propellers shared the same baseline geometry, including, but not limited to, their diameter, chord distribution, and airfoil profile. The only difference between the two propellers was the addition of surface roughness, applied to the top (suction) side of the rough propeller's blades. The propellers were operated within the low-$Re$ regime, and the simulations were run for sufficient time to achieve converged values of thrust and torque.

\subsection{Flow Solver Information}
\label{ssec:flowSolverInfo}

The flow physics of the rough and smooth propellers were uncovered and compared against each other through \textit{Charles}, a cell-centered, unstructured grid LES flow solver, produced by Cadence Design Systems. The code deployed a second order accurate and energy-conserving method to solve the compressible Navier-Stokes equations. Time integration was performed using the explicit, third order Runge-Kutta scheme. For the calculations presented, no subgrid-scale model was employed, 
because the Reynolds number was relatively low to 
expect little to no turbulence and sufficiently fine grids were used. \textit{Charles} has been deployed for use of high-fidelity calculations with complex geometries, such as wall-modeled LES of an airplane in a landing configuration~\cite{goc20201wmles}, prediction of supersonic jet performance with unstructured LES~\cite{bres2017unstructured}, and most recently for wall-modeled LES of the NASA juncture flow~\cite{lozanoduran2022performance}. {Two recent studies on fully rotating turbomachinery from Powers and Gilbert~\cite{powers2023unsteady} and Jain~\etalnp\cite{jain2020massively} have deployed \textit{Charles} with a rotating reference frame. The Navier-Stokes equations with the added centrifugal and Coriolis forces are the same for the preceding studies and the present study.} The entire code was written in C++, and utilized both domain decomposition and Message Passing Interface (MPI) for parallelization. 

Voronoi cells generated by a hexagonal close-packed point-seeding method were used for iterative grid generation and refinement within \textit{Charles}. These cells were locally isotropic. For formulating the mesh, the far-field grid spacing $\Delta$~[\si{\meter}] was first set. This far-field value was used to determine the size of subsequent levels of refinement. With each iteration of the mesh, either the far-field $\Delta$ or the number of desired refinement levels could be changed. Lloyd's algorithm, which is able to partition Voronoi grids into uniformly sized cells, was deployed in a smoothing process to ensure adequate blending between refinement levels. This iterative algorithm was completed ten times at each refinement layer.  Further information on the grid generation method is outlined in~\cite{lozanoduran2022performance} 
 
\subsection{Domain Setup, Meshing, and Boundary Conditions}
\label{ssec:bcs_domainSetup}

\begin{figure}[h]
    \centering
    \includegraphics[width=0.85\textwidth]{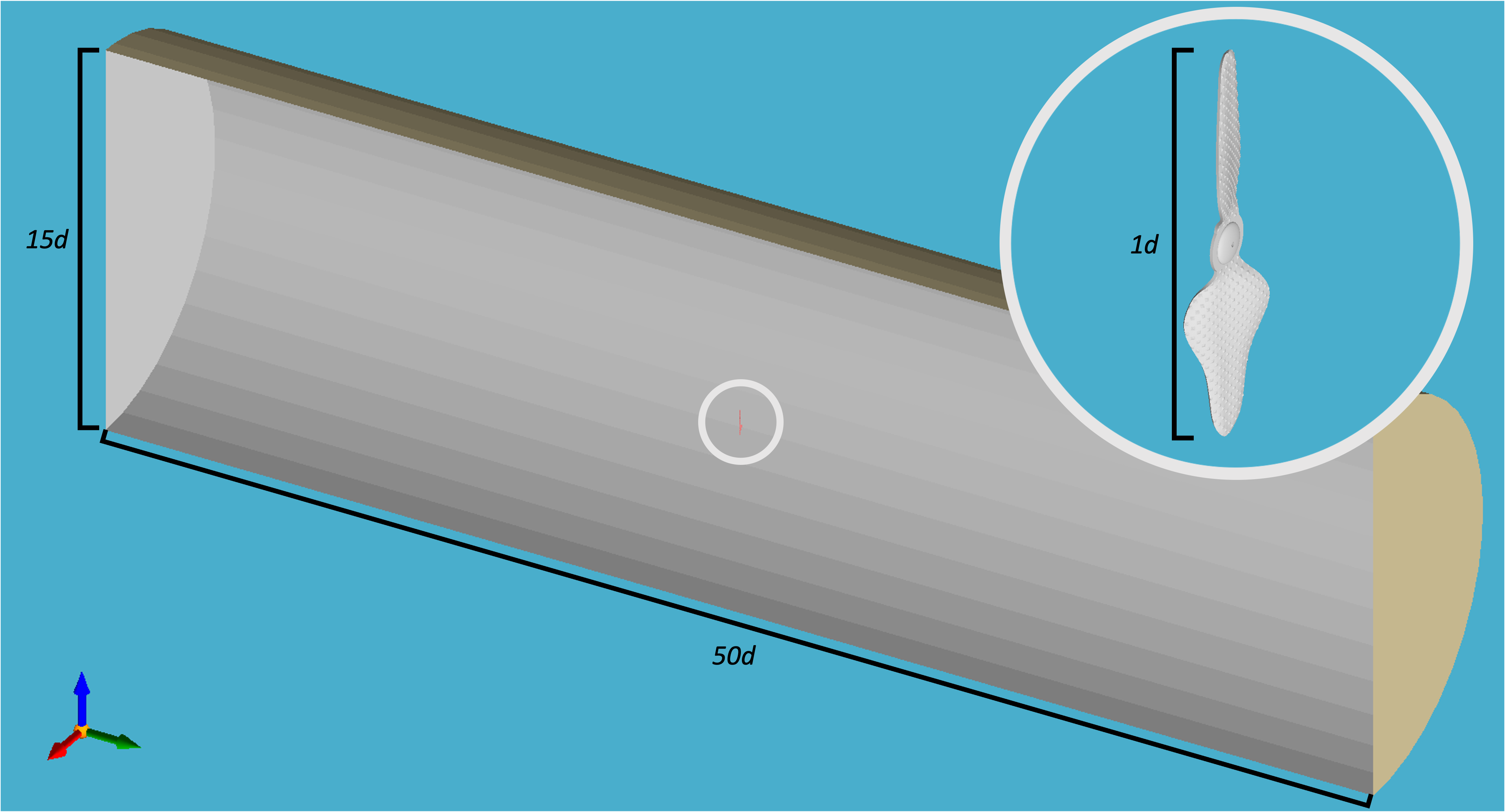}
    \caption{Cut section view of the computational domain deployed in the study, including an expanded view of the propeller.}
    \label{fig:prop_domain}
\end{figure}

The compressible Navier-Stokes equations were solved in a rotating reference frame, within a cylindrical domain, to simulate propeller rotation beginning from rest. The rotation was fixed at a constant angular velocity $\Omega$~[\si{\radian\per\second}], held about the $y$-axis, in order to obtain a desired value of the $Re_c$. For this study, $Re_c = 6,500$ was chosen since it resides well within the low-$Re$ regime, and it remains relevant to current and future MAV design. The propeller was placed at the center of the enclosed domain, as shown in Fig.~\ref{fig:prop_domain}. Symmetry boundary conditions were applied to the top, bottom, and side walls of the domain, while a no-slip, isothermal wall boundary condition was used on the propeller surface. The domain had a length of $50d$ and a diameter of $15d$; an exploratory study was conducted prior to the investigation to determine adequate domain sizing and mesh independence, as discussed in Sec.~\ref{sssec:dg_study}.  

\subsubsection{Domain and Grid Independence Studies}
\label{sssec:dg_study}

We carried out a preliminary study to ensure the enclosed domain size did not interfere with force measurements. Multiple domain sizes were tested with the smooth propeller, and time-averaged values of $T$ were recorded to determine domain influence on forces. For all domains, the mesh structure and near-blade resolution were unchanged. A constant $\Omega$ value was used to keep the same $Re_c$ for each case. Two domain lengths were tested to ensure blockage effects do not interfere with force measurements. The longer of the two was $100d$ in length, and it was compared against a domain with a length of $50d$. For both domains, the diameter was held constant at $15d$. This chosen diameter was more than double the size of that used in a similar calculation conducted by Kumar and Mahesh~\cite{kumar2017large}, who performed an LES study on marine propeller wake instabilities. The resulting difference in thrust force after approximately 10 rotations was less than 1\%. Therefore, the shorter $50d$ domain was selected for the numerical study to reduce overall grid count. 

To capture the effects of the roughness, adequate near-wall cell spacing was required. A grid independence study was carried out to ensure the convergence of force measurements for the calculation. This independence study was conducted for both propellers, to ensure convergence in the presence of surface roughness, and without roughness. The domain size was kept constant with the length and radius as found in the domain sizing study. Multiple iterations of meshes with varying refinement near the blade, as well as the mesh structure, were studied. Between the finest mesh deployed in the grid study, and the next finest case, the difference in observed force values was less than 1\%. Thus, the finest mesh from the grid independence study was selected for the numerical investigation. 

\begin{figure}[hbt!]
    \centering
    \begin{subfigure}[t]{0.80\textwidth}
        \centering
        \includegraphics[width=\textwidth]{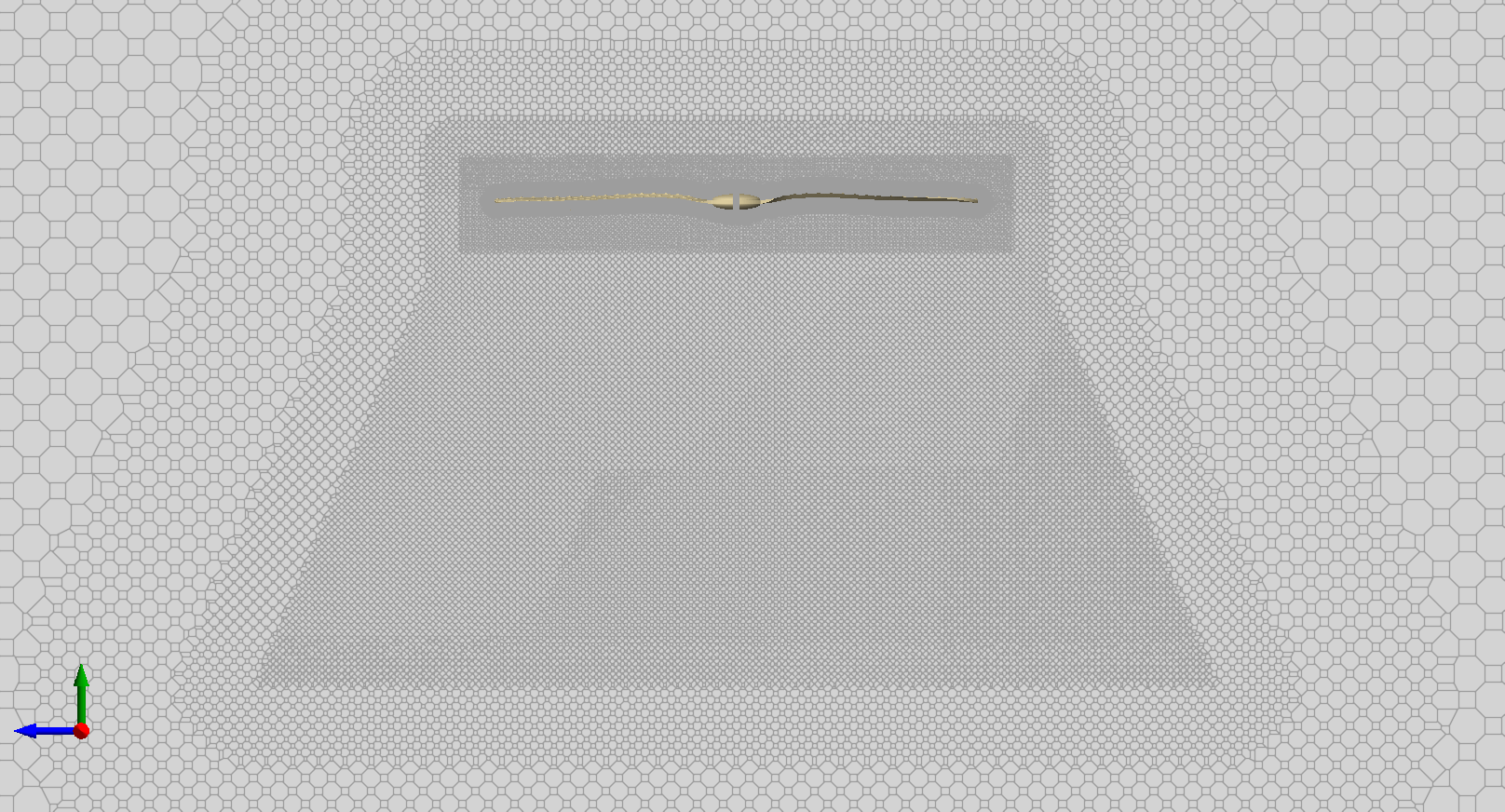}
        \caption{The propeller wake region.}
        \label{fig:XPlane_WakeRegion}
    \end{subfigure}
    \hfill
    \begin{subfigure}[t]{0.80\textwidth}
        \centering
        \includegraphics[width=\textwidth]{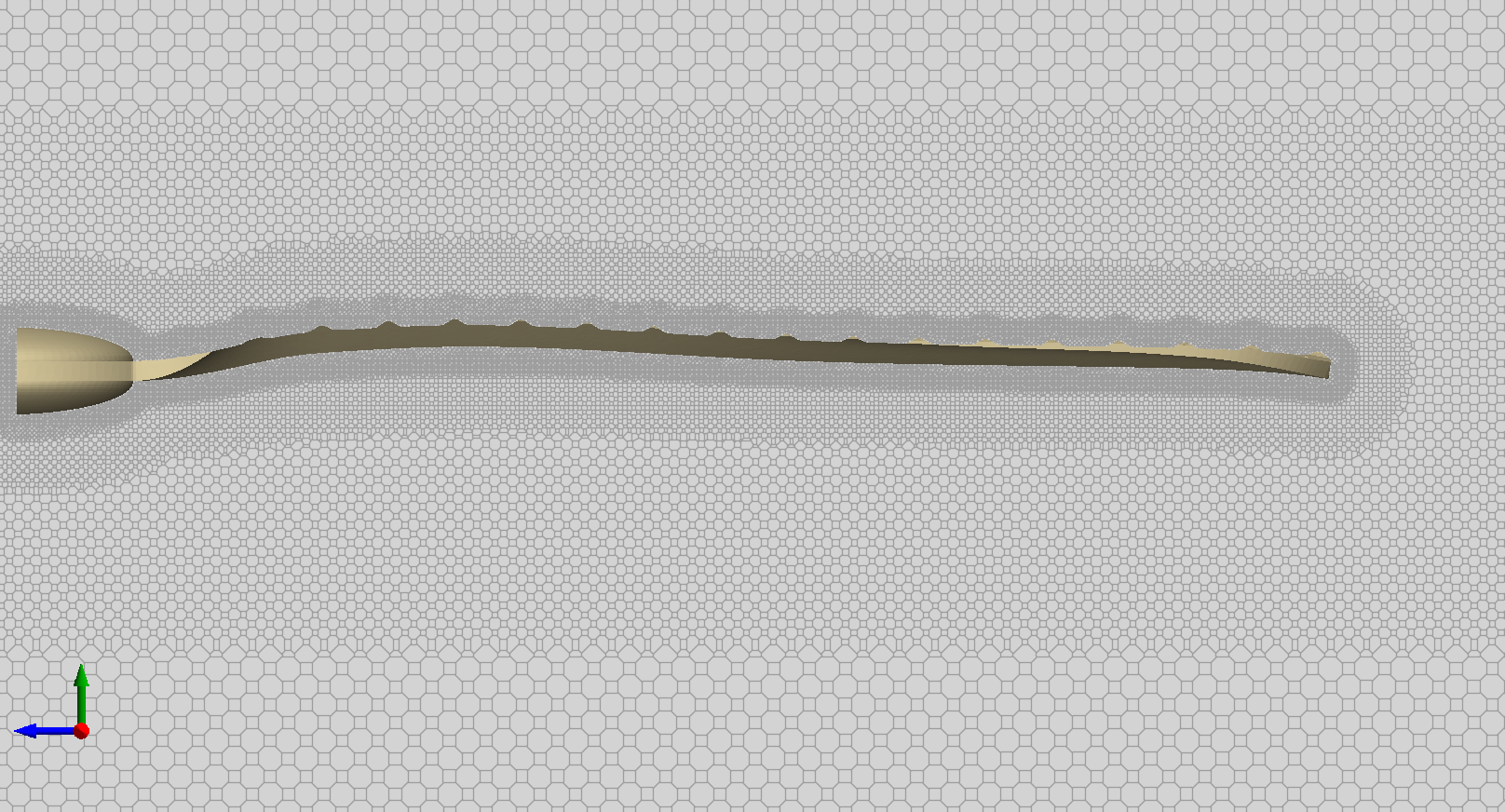}
        \caption{Close-up of the near-blade mesh.}
        \label{fig:blade_closeup}
    \end{subfigure}
    \caption{Mesh refinement near the propeller}
    \label{fig:mesh}
\end{figure}

{Beginning with the largest grid spacing in the far field, each subsequent refinement level is half the size of the previous spacing, until $\Delta_{min}$ is reached.} For the mesh, the minimum grid spacing, which corresponds to the cell size in the region closest to the propeller surface, was localized to a thin layer ten cells thick around the propeller, {with} $\Delta_{min}=8.45\times10^{-4}d$. {Due to the cells being isotropic, 
the near-wall grid spacing in viscous wall units  were 
$\Delta x^+ = \Delta y^+ = \Delta z^+ \equiv \Delta^+$. Across the 75\% blade-station, the  near-wall grid spacing averaged over the chord was 
$\Delta^+ \approx 2.85$, and similarly the average value from root-to-tip for average was $\Delta^+ \approx 2.70$. } Each roughness element was resolved by approximately three mesh elements in the wall-normal direction. The next finest cells were confined to a disk centered around the propeller with diameter $1.075d$, extending $0.1d$ in front and behind the propeller, {and would be twice the size $\Delta_{min}$}. Further refinement was also applied to the wake $1d$ downstream of the propeller, as well as in the center of the domain $12.5d$ upstream and $15d$ downstream. Figure~\ref{fig:XPlane_WakeRegion} provides a visual of the wake region mesh structure, and Fig.~\ref{fig:blade_closeup} gives a close-up view of the mesh refinement surrounding the blade.

\section{Results and Discussion}
\label{sec:results}

\subsection{Numerical Results}
\label{ssec:numResults}

\subsubsection{Effect of Surface Roughness on Propeller Thrust}
\label{sssec:thrustEffects}

We first compared the thrust produced by the rough propeller to that from the baseline smooth propeller. The rough propeller showed an improvement in thrust force values, resulting in $1.95\%$ higher thrust at the same $Re_c$, with the percent difference calculated using 
\begin{equation}
    \% \text{Diff} = \frac{\Bar{X}_{rough} - \Bar{X}_{smooth}}{\Bar{X}_{smooth}} \times 100. 
    \label{perDiff}
\end{equation} 
The roughness was also found to decrease $C_P$  by $4.92\%$. Values of $\Bar{C}_T$ and $\Bar{C}_P$ are presented in Table~\ref{tab:thrustForces} for both the rough propeller and smooth (baseline) propeller. {Prior studies using bio-inspired surface roughness to reduce drag produce a wide range of results, from as little as 2.8\% to as high as 61\%, with the majority yielding $\sim O(1)$ percent improvements ~\cite{liu2020brief}. Considering one of the most limiting factors for MAVs is allowable battery size and energy storage~\cite{yue2021nearly}, improvements in $C_T$ and $C_P$ of any magnitude should be considered significant.} We also note again that the propeller and its associated roughness are not optimized for this flow condition. Our principle aim in this work was to discern the underlying flow physics when roughness is present on a propeller at low $Re$.  Future efforts to co-optimize the propeller and roughness may yield additional  performance improvements.

{The thrust found for the baseline propeller provides a reasonable estimation based on prior work for this propeller design. Kroo~\etal~\cite{kroo2001mesicopter} ran simulations for a four-blade version at 48,000 RPM ($Re \approx 10,000$), yielding a predicted $4~\si{\gram}$ of thrust. Our two-blade propeller operating at a lower 13,260 RPM predicts an approximate $1.8~\si{\gram}$ of thrust, giving confidence to the results of the simulation. Reviewing the components of 
the thrust force $T$,} the rough propeller showed greater pressure forces, {compared to the baseline propeller, with viscous forces being negligible.} The roughness elements present on the suction side of the blade may be contributing to a reduction of form drag. Zhou and Wang~\cite{zhou2012effects} similarly found that leading-edge roughness elements on an airfoil at $Re = 60,000$ decreased overall drag by decreasing pressure drag associated with laminar separation bubbles. Choi, Jeon, and Choi~\cite{choi2006mechanism} investigated sphere drag reduction by dimples and found the roughness provides more momentum to the flow through a separation and reattachment mechanism, allowing the flow to overcome any adverse pressure gradients on the surface. The roughness potentially displays a similar mechanism, as the pressure is observed to decrease as the flow encounters each element.  

\begin{table}[hbt!]
\caption{\label{tab:thrustForces} {Comparison of $\Bar{C}_T$ and $\Bar{C}_P$ between the rough and smooth propellers.}} 
\centering
\begin{tabular}{lrcrc}
\hline
\hline
Propeller & $\Bar{C}_T$ & \% Diff. of $\Bar{C}_T$ & $\Bar{C}_P$ & \% Diff. of $\Bar{C}_P$ \\ 
\hline
Smooth & {0.1175} & -      & {0.07757} & - \\
Rough  & {0.1198} & 1.95\% & {0.07375} & 4.92\% \\
\hline
\hline
\end{tabular}
\end{table}

\begin{figure}[hbt!]
    \centering
    \begin{subfigure}[t]{0.75\textwidth}
        \centering
        \includegraphics[width=0.7\textwidth]{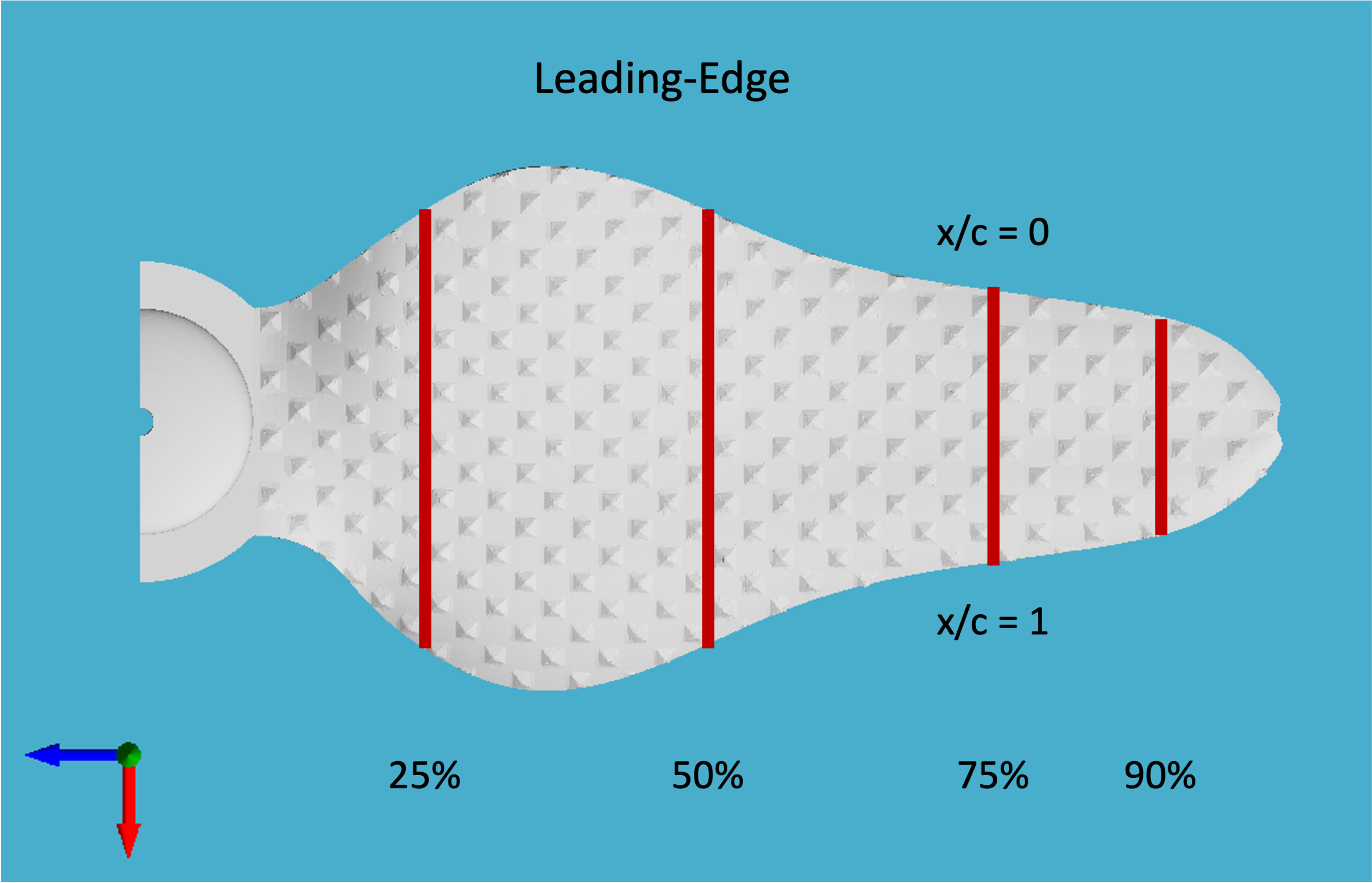}
    \caption{Locations of the 25\%, 50\%, 75\%, and 90\% blade-stations used for capturing pressure along the suction and pressure sides of the blade.}
    \label{fig:bladestations}
    \end{subfigure}
    \hfill
    \begin{subfigure}[t]{0.9\textwidth}
        \centering
        \includegraphics[width=0.9\textwidth]{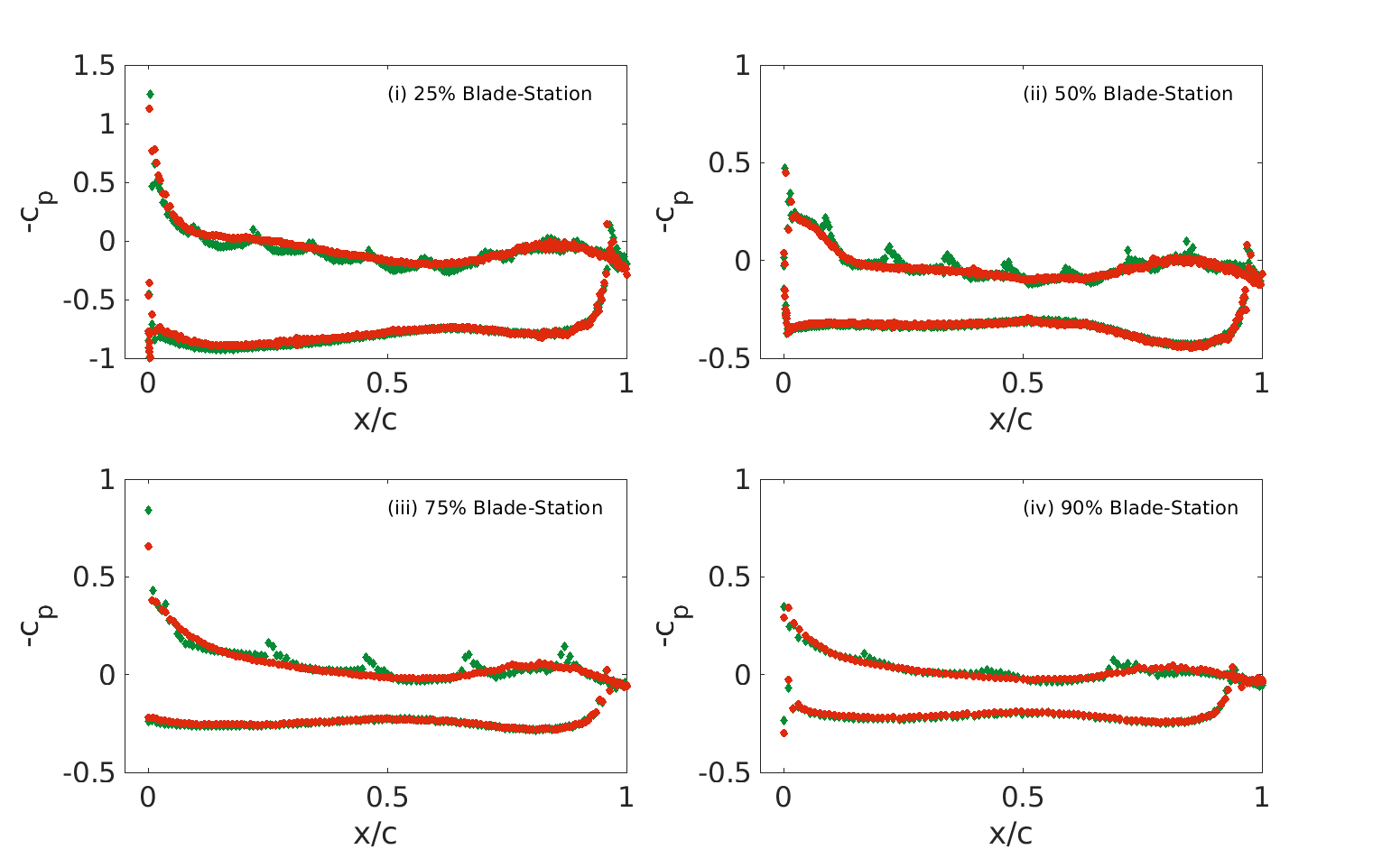}
    \caption{$-c_p$ plot comparing the rough (green diamonds) and smooth (red circles) propellers at the (i) 25\% (ii) 50\% (iii) 75\% and (iv) 90\% blade-stations.}
    \label{fig:cp}
    \end{subfigure}
    \caption{
    Sampling locations and 
    respective chordwise surface pressure distributions at select spanwise locations. }
    \label{fig:cp_w_loc}
\end{figure}

The pressure coefficient on the suction and pressure sides of the propeller was reviewed to analyze the impact of surface roughness and changes in thrust. Wall pressure was sampled at the 25\%, 50\%, 75\%, and 90\% blade-stations, as shown in Fig.~\ref{fig:bladestations}. These blade-stations show the progression of $c_p$ from the root to the tip, and adequately compare the roughness effects. Figure~\ref{fig:cp} shows the $c_p$ distribution for both propellers at each of the aforementioned blade-stations. A clear difference between the two is observed.

Looking first at the baseline results, represented by red circles in Fig.~\ref{fig:cp}, for all blade-station locations, the 
suction peak with negative $c_p$ begins near  the leading-edge, which is followed by increase in pressure as the flow moves downstream. The pressure is observed to decrease slightly from near $x/c = 0.6$, plateaus near $x/c = 0.8$, and then finally increases  a bit towards the trailing edge. 
Although $c_p$ appears to plateau near the 
$0.8c$ location (often indicative of flow separation~\cite{arena1980laminar}), observation of the wall shear stress and the flow direction close to the blade surface do not indicate separation is occurring.    

\begin{figure}[hbt!]
    \centering
    \begin{subfigure}[t]{0.7\textwidth}
        \centering
        \includegraphics[width=\textwidth]{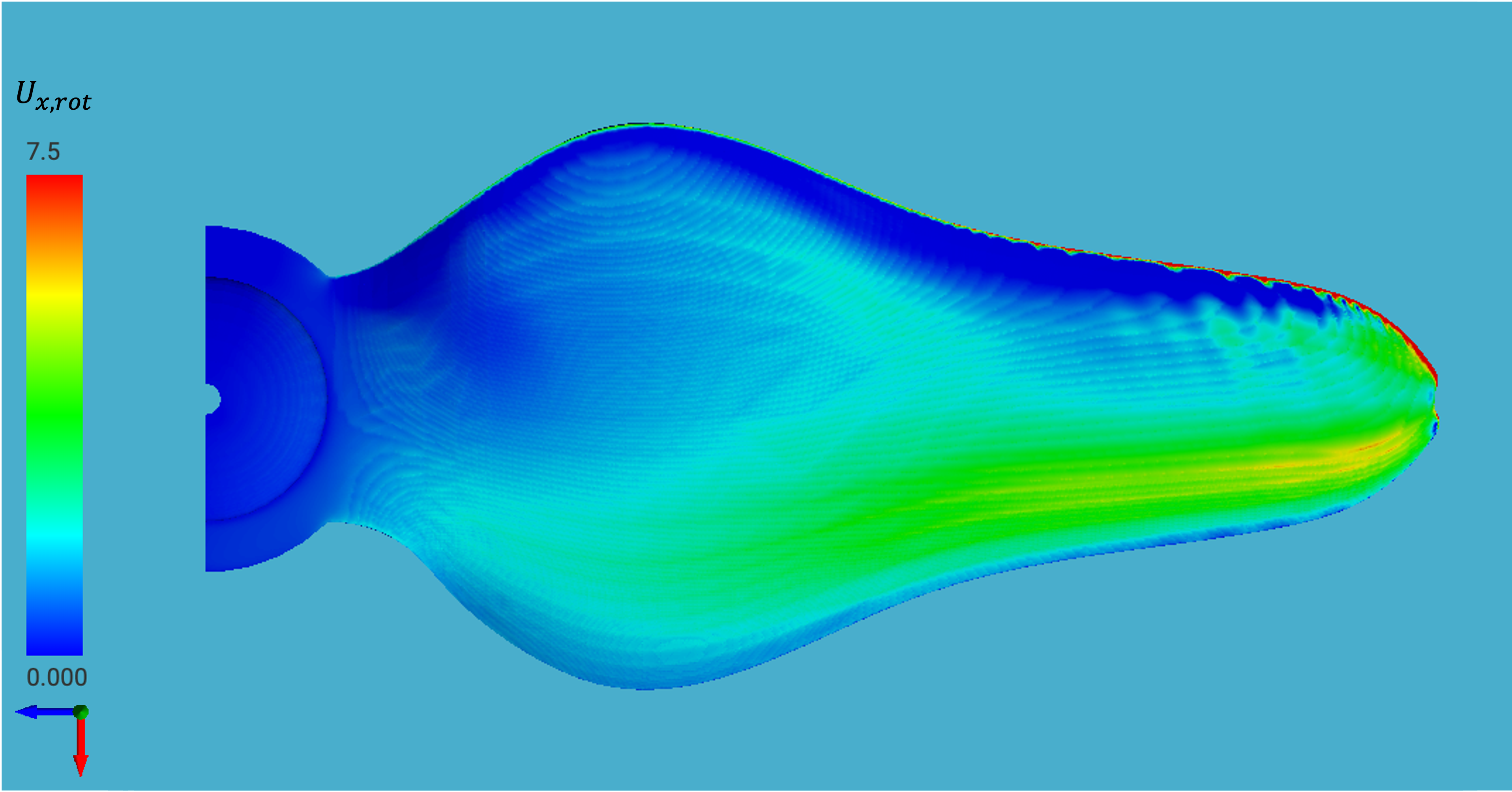}
        \caption{Smooth propeller.}
        \label{fig:sp_projurot}
    \end{subfigure}
    \hfill
    \begin{subfigure}[t]{0.7\textwidth}
        \centering
        \includegraphics[width=\textwidth]{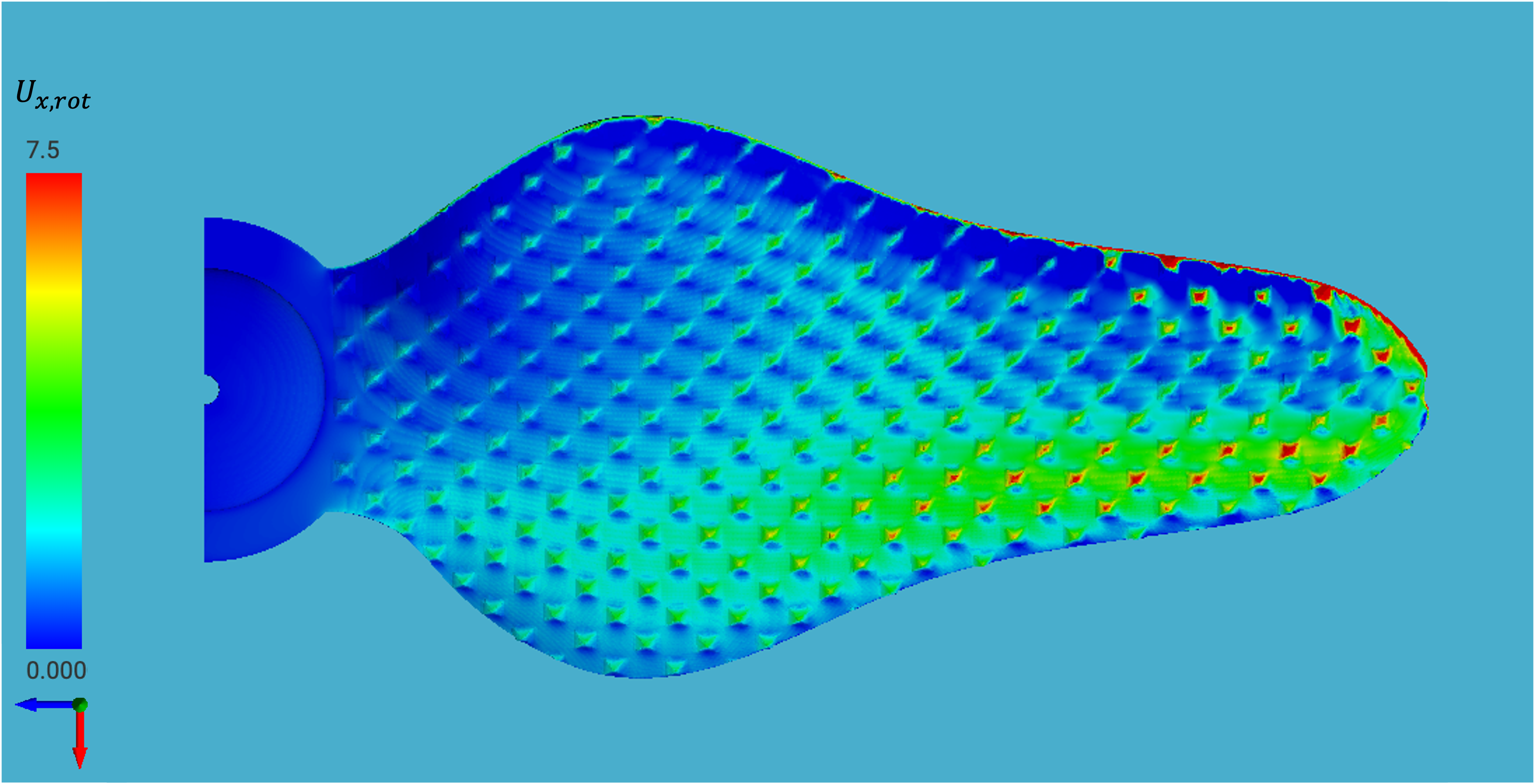}
        \caption{Rough propeller.}
        \label{fig:rp_projurot}
    \end{subfigure}
    \caption{{
    Contour of the chordwise surface velocity  
    displaying local increase of flow speed at the 
    apexes of the roughness elements. The velocity here is measured relative to the frame rotating with the blade. The velocity at the center of the wall-adjacent cells were projected on to the blade surface. 
    The leading-edge is as indicated in Fig.~\ref{fig:bladestations}.}}
    \label{fig:projUxROT}
\end{figure}

The rough propeller, given by green diamonds in Fig.~\ref{fig:cp}, experiences higher suction peaks, compared to the baseline, at each blade-station. The 
pressure starts to increase downstream of the peak as it did with the smooth propeller. However,  characteristic oscillation in the pressure is clearly observed as the flow encounters roughness elements repeatedly downstream. 
A study by Ferreira and Ganapathisubramani~\cite{ferreira2020piv}, who examined canopy-like cube roughness, found that regions of low pressure may form on the windward side of roughness elements, and then transition to high pressure regions as the flow moves over top of the roughness elements, similar to the phenomena seen with the propeller roughness elements in the present study. {These 
local pressure peaks coincide with the apexes of the roughness elements where the flow velocity is locally higher than its surrounding, (see comparison in Figure~\ref{fig:projUxROT}). 

} 

Furthermore, research on vortex generators has shown that the addition of perturbations provides similar tendencies in altering pressure distributions across airfoils. Work to investigate the ability of vortex generators to delay dynamic stall by Heine~\etal~\cite{heine2013dynamic} showed that by placing roughness elements near the leading-edge of an airfoil led to increased suction peaks of pressure coefficients, and enhanced lift capabilities at higher angles of attack compared to baseline airfoils. In addition, De~Tavernier~\etal~\cite{tavernier2021controlling} found that adding a strip of vortex generators at mid-chord locations ($0.2c$, $0.3c$, and $0.4c$) on a wind turbine airfoil enhanced the flow's pressure distribution and prevented separation at higher angles of attack than when placed closer to the leading-edge. 
Overall, in the present study, roughness elements reduces the pressure on the suction side primarily through augmentation of the suction pressure peak near the leading edge  as well as introduction of new local pressure minima, leading to increases in thrust.

\subsubsection{Influence of Surface Roughness on Vortex Structures}
\label{sssec:vortexStructures}

Vortex structures near the leading-edge, trailing-edge, and at the tip of the propeller blade were examined next. Vortices have the ability to adversely impact propellers and wings if they detach or shed, reducing overall lift capabilities, but they may also enhance lift if they remain attached and stable~\cite{jardin2021empirical,ford2013lift}. These structures were found to be affected by the inclusion of surface roughness on the propeller blade. In particular, the vortices near the leading-edge and tip show differences when comparing the rough propeller to the baseline. 
To aid in analyzing the vortex structures, iso-surfaces of $Q_{II}$ (the second invariant of the velocity gradient tensor), which indicate regions of 
low straining and high swirling motions~\cite{hunt1988eddies}, are visualized. These iso-surfaces are colored by {instantaneous values of} the pressure coefficient, magnitude of absolute velocity, as well as magnitude of absolute vorticity, to further dissect the flow physics. {
Note that these vortex structures remained unchanged on all grids considered. } 

\begin{figure}[hbt!]
    \centering
    \begin{subfigure}[t]{0.75\textwidth}
        \centering
        \includegraphics[width=\textwidth]{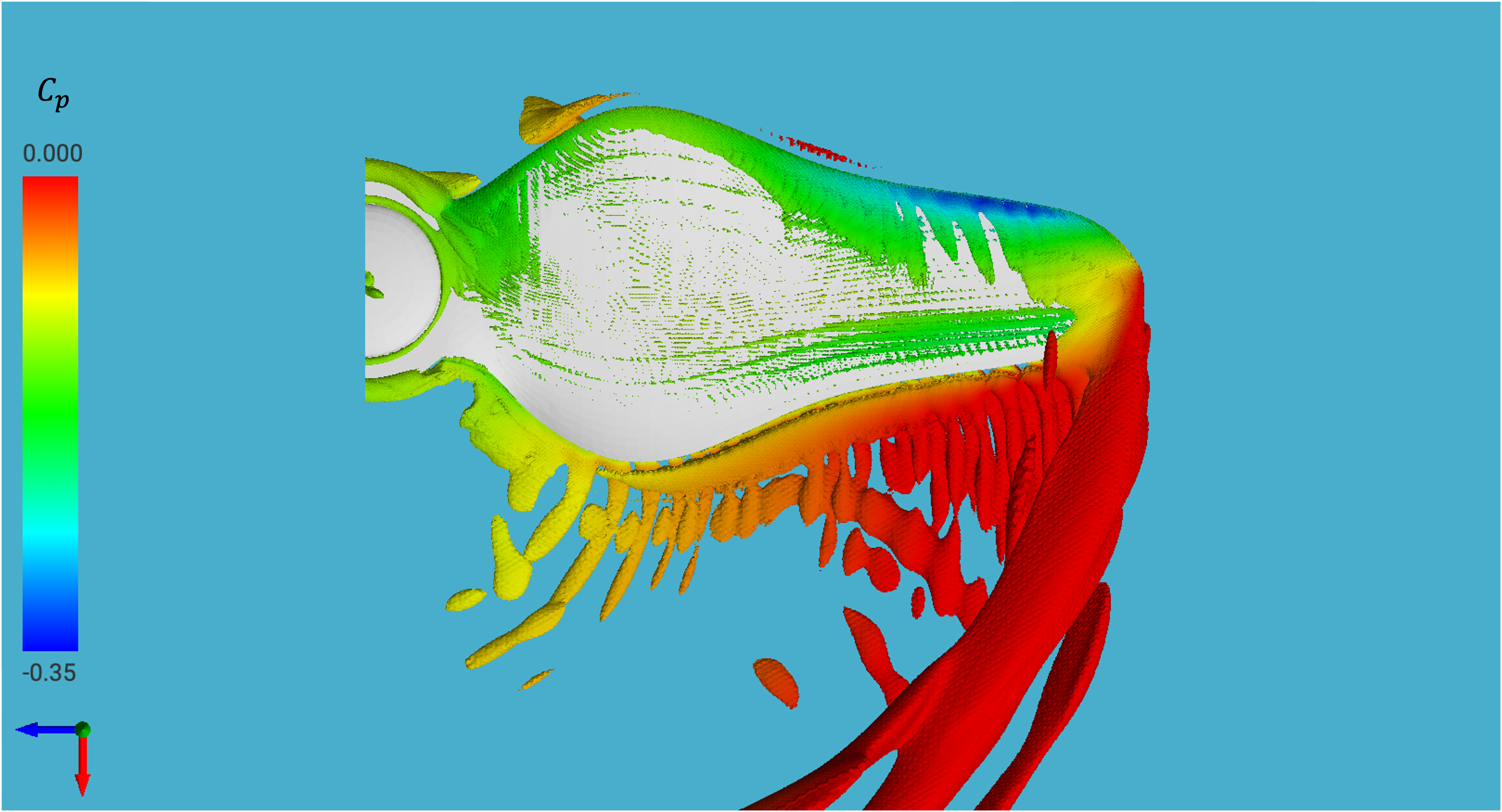}
        \caption{Smooth propeller.}
        \label{fig:sp_q2_cp}
    \end{subfigure}
    \hfill
    \begin{subfigure}[t]{0.75\textwidth}
        \centering
        \includegraphics[width=\textwidth]{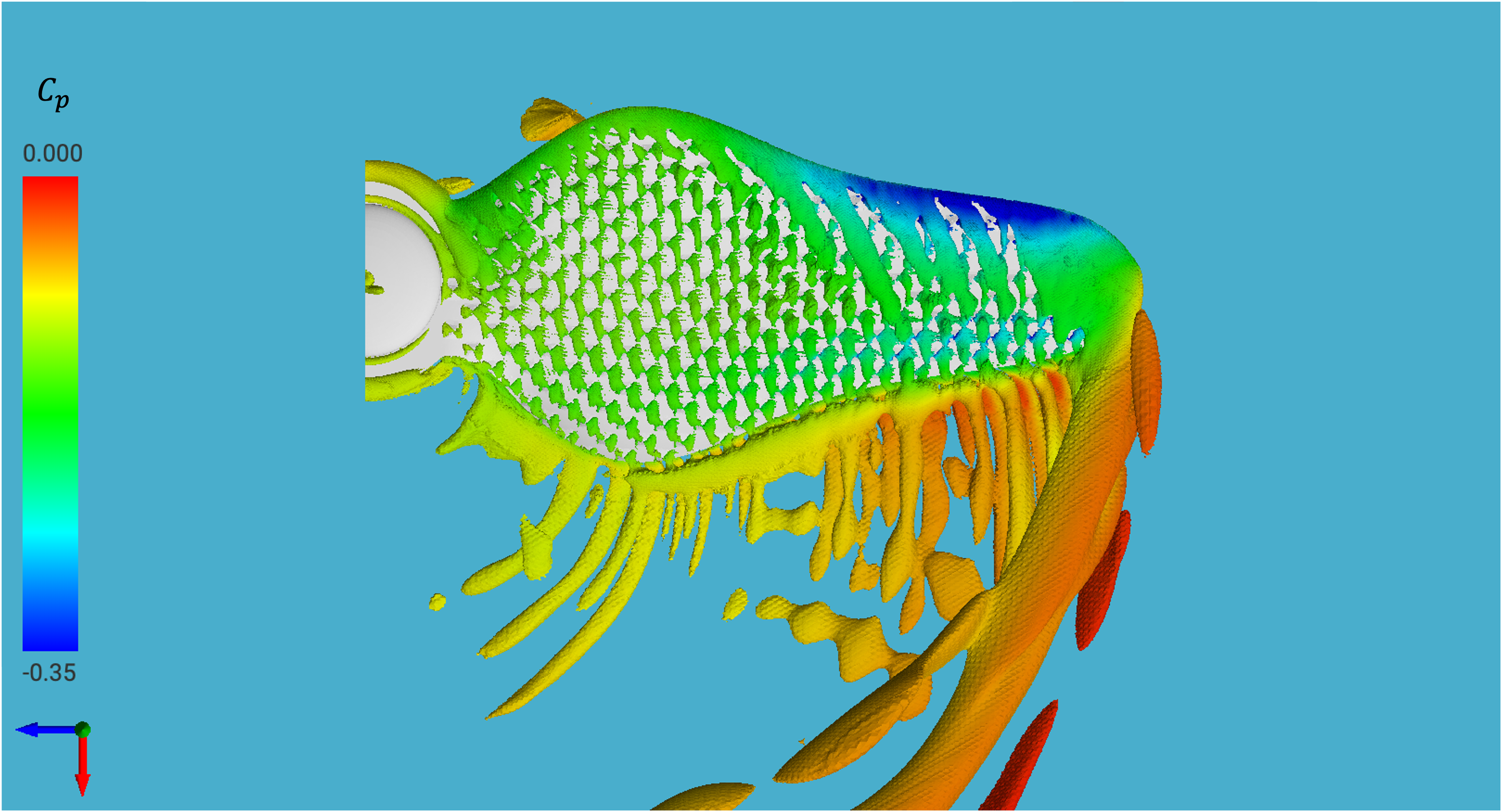}
        \caption{Rough propeller.}
        \label{fig:rp_q2_cp}
    \end{subfigure}
    \caption{    Isosurfaces of $Q_{II}$  (the second invariant of the velocity gradient tensor) colored by the pressure coefficient. 
    (blade rotating CCW)}
    \label{fig:q2Cp}
\end{figure}

\begin{figure}[hbt!]
    \centering
    \begin{subfigure}[t]{0.75\textwidth}
        \centering
        \includegraphics[width=\textwidth]{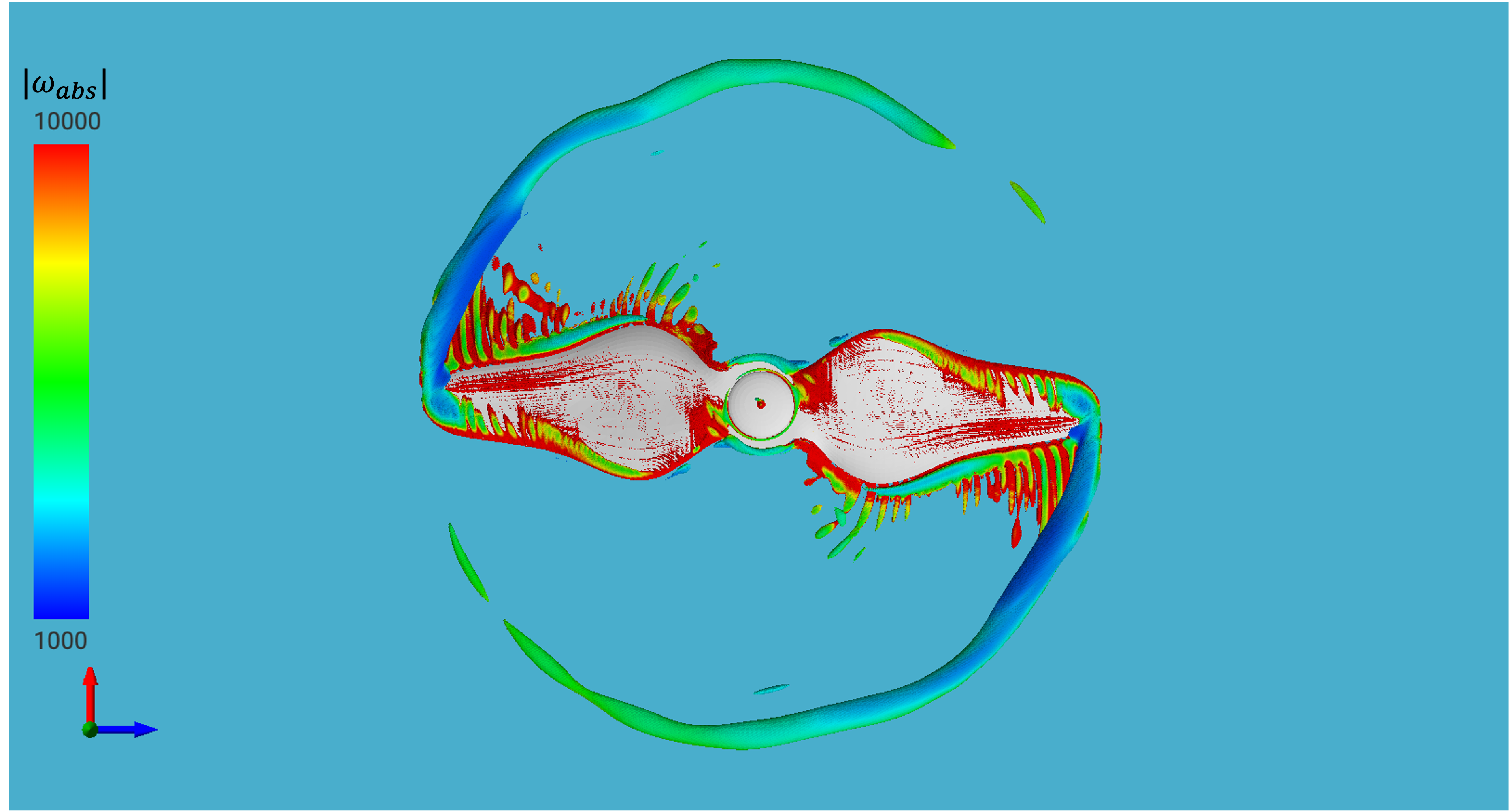}
        \caption{Smooth propeller.}
        \label{fig:sp_q2_vortabs}
    \end{subfigure}
    \hfill
    \begin{subfigure}[t]{0.75\textwidth}
        \centering
        \includegraphics[width=\textwidth]{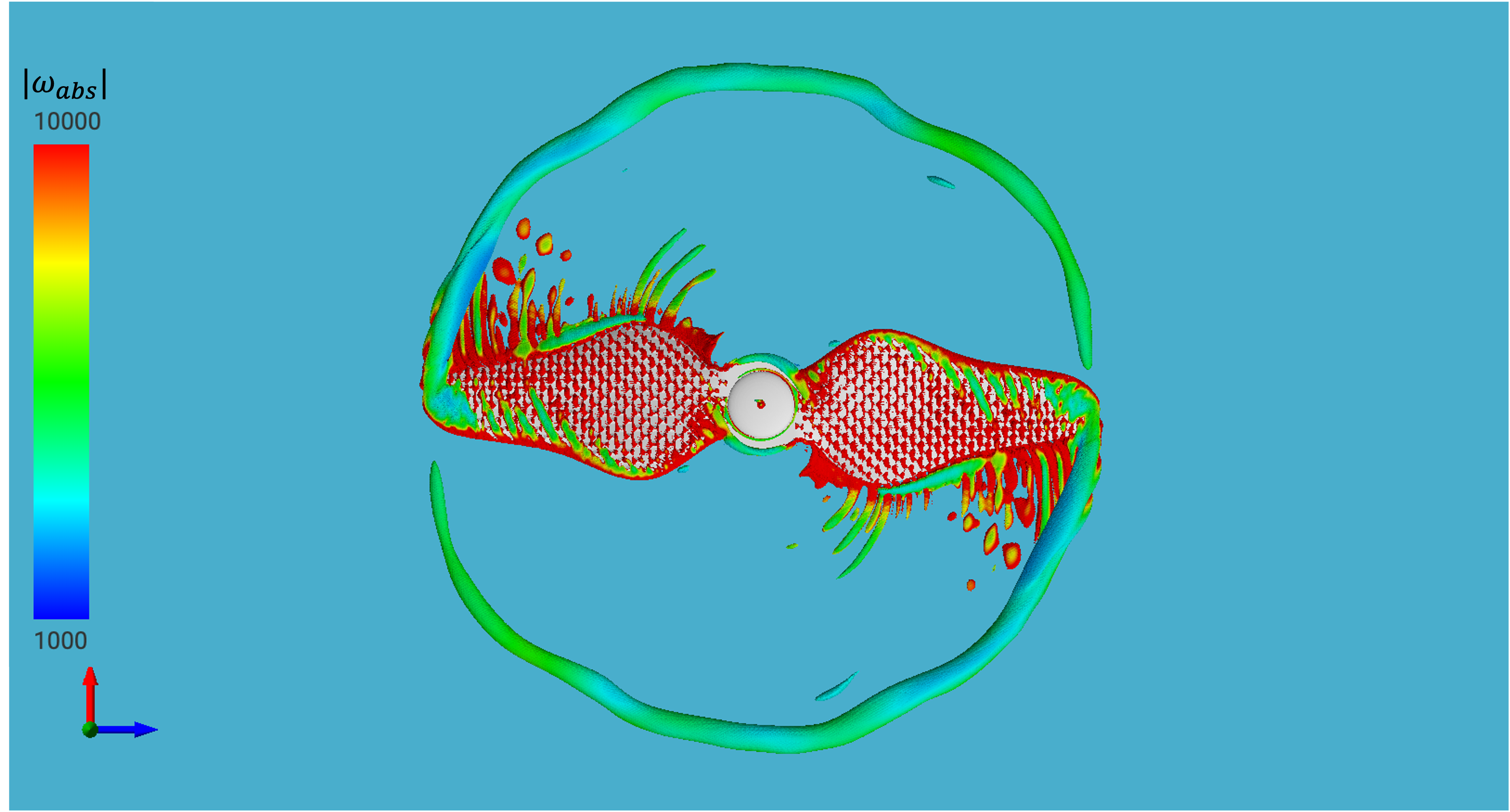}
        \caption{Rough propeller.}
        \label{fig:rp_q2_vortabs}
    \end{subfigure}
    \caption{Isosurfaces of $Q_{II}$  (the second invariant of the velocity gradient tensor) colored by  the vorticity magnitude 
    ($|\omega_{abs}|$) (blade rotating CCW).}
    \label{fig:q2Vortabs}
\end{figure}

{In general, lower pressure regions correspond to stronger vortex cores \cite{hunt1988eddies}.} In the case of the smooth propeller, it can be seen from Fig.~\ref{fig:sp_q2_cp} that there is a low pressure region along the leading-edge of the blade, which is characteristic of the LEV \cite{jardin2017Coriolis}. Few iso-structures exist in the mid-chord; however, there exists above the trailing-edge a long streak moving out towards the tip, indicating the presence of vortex structures. At the trailing-edge there are some structures that ultimately merge with the tip vortex, which contains regions of high pressure. {For a rotating body, Jardin~\cite{jardin2017Coriolis} found the Coriolis forces associated with rotation allow for a stable LEV to be maintained, and then merge with the tip vortex. This is found to occur with the baseline propeller, and the length and strength of the resulting tip vortex is shown in Fig.~\ref{fig:sp_q2_vortabs}.} 

The rough propeller, shown in Fig.~\ref{fig:rp_q2_cp}, displays a {physically} larger low pressure region at the leading-edge. This alludes to a stronger LEV with the roughness present (also evidenced by the extension of the suction peak region) when compared to the smooth propeller. As found by Nabawy and Crowther~\cite{nabawy2014quasi}, a stable LEV that remains attached to the wing may augment lift forces. Both propellers show stable LEVs, yet the rough propeller {maintains 
more negative $c_p$, and hence enhanced suction peak values, further towards the tip}. This is indicative of a more stable, and stronger, LEV on the rough propeller, compared to the baseline propeller. Further evidence of the enhanced LEV is seen by the strengthened tip vortex, shed by the rough propeller. Figure~\ref{fig:rp_q2_vortabs} {provides a visual of the tip vortex, which is shown to extend further into the propeller wake, as well as having larger magnitudes of vorticity, indicating stronger swirling within.}

\begin{figure}[hbt!]
    \centering
    \begin{subfigure}[t]{0.75\textwidth}
        \centering
        \includegraphics[width=\textwidth]{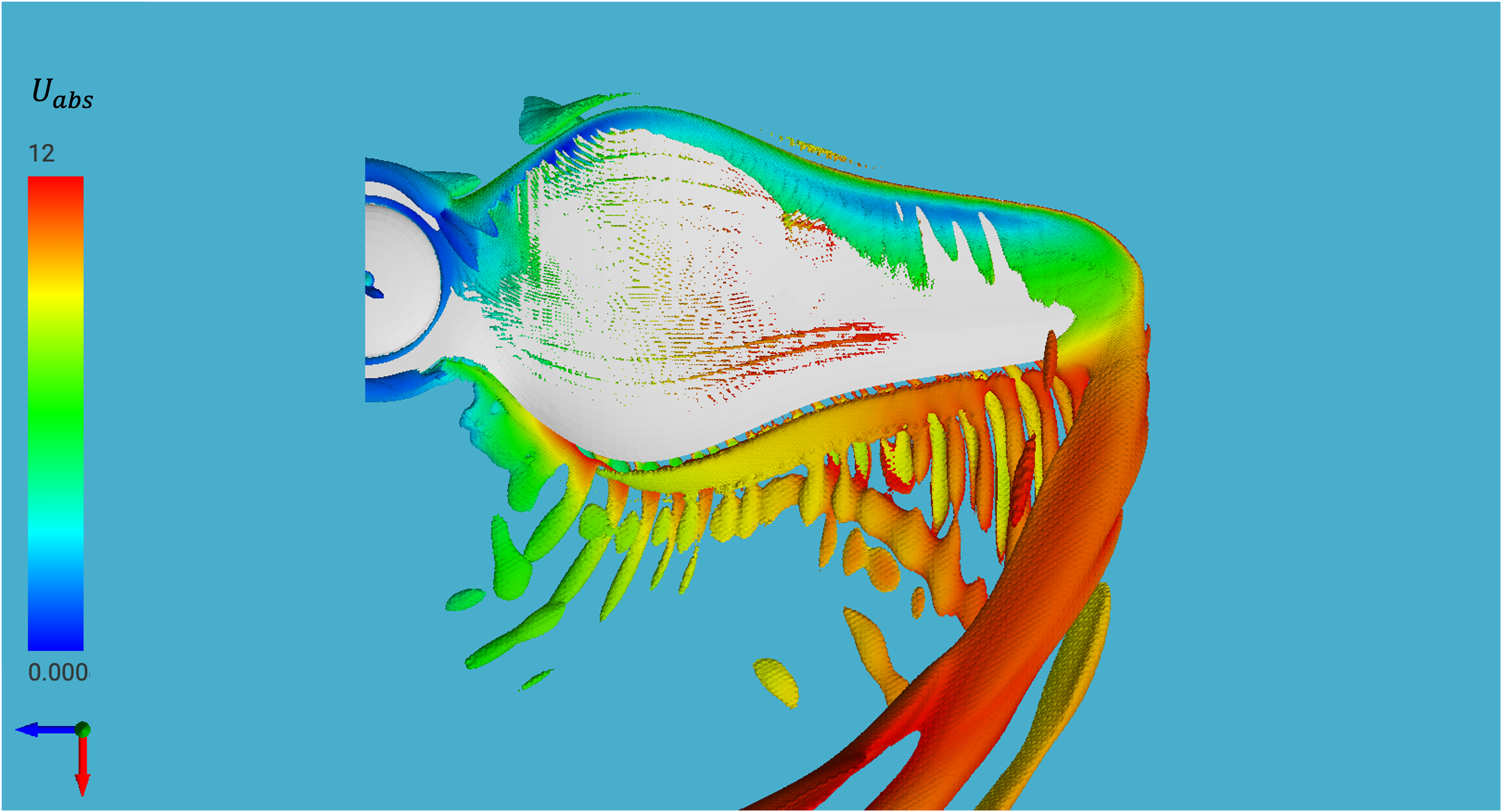}
        \caption{Smooth propeller.}
        \label{fig:sp_q2_uabs}
    \end{subfigure}
    \hfill
    \begin{subfigure}[t]{0.75\textwidth}
        \centering
        \includegraphics[width=\textwidth]{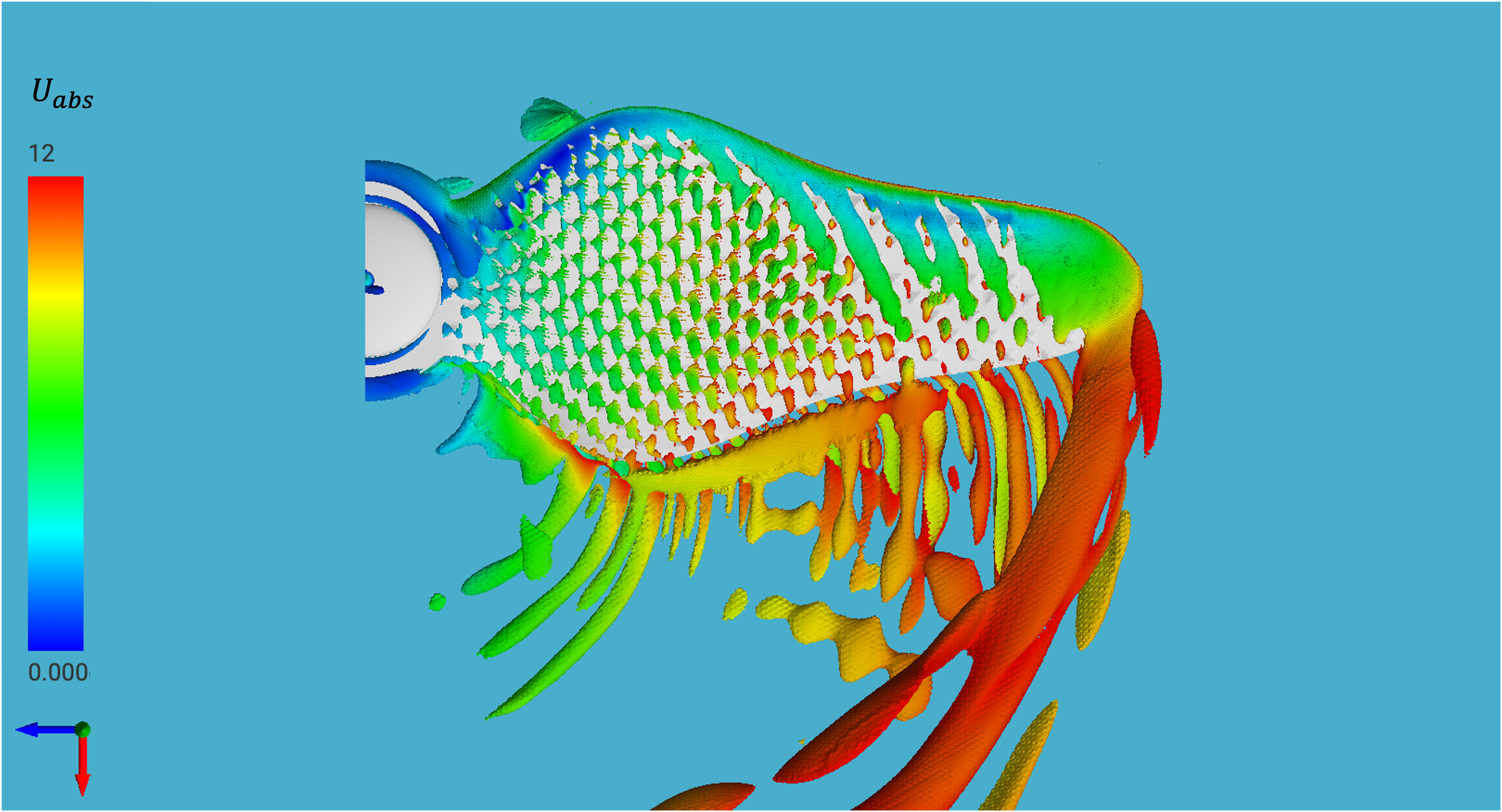}
        \caption{Rough propeller.}
        \label{fig:rp_q2_uabs}
    \end{subfigure}
    \caption{Isosurfaces of $Q_{II}$  (the second invariant of the velocity gradient tensor) 
    colored by the velocity magnitude ($|U_{abs}|$). }
    \label{fig:q2Uabs}
\end{figure}

Elsewhere on the blade, due to the presence of the roughness, there are iso-structures located across the mid-chord region that are not seen on the smooth propeller. At the trailing-edge and near the tip, the structures appear to be similar to those on the smooth propeller. However, these surfaces are at a noticeably lower $c_p$ in the case of the rough propeller. Figure~\ref{fig:q2Uabs} provides a view of the structures colored by magnitude of the absolute velocity for both propellers. With the rough propeller, there are some lower velocity regions within the spacing between elements, as well as indications of higher velocity flow over top the roughness, further validating the presence of low $c_p$ regions over the top of the roughness elements. Note that the roughness allows for flow acceleration over areas of the propeller surface where the smooth propeller sees the flow decelerate. 

\subsection{Experimental Results}
\label{ssec:expt}

The numerical results above suggest that adding roughness to a propeller can improve its performance. To corroborate our numerical findings, we 3D-printed these propellers at a resolution of 0.25~\si{\micro\meter} (Formlabs Form3 printer) and used a thrust stand, designed similarly to that used by Deters~\etal~\cite{deters2008static, deters2014reynolds, Deters2018-4122, Jordan2020-2595} (note also that used by Kruyt~\etal~2014~\cite{Kruyt2014-20140585}, and see also the work of Cha~\etal~\cite{Cha2020-1127}), to evaluate their thrust and drag coefficients. We will present only the most important details of these experiments here; a complete description is available in the Supplemental Materials. Briefly, our thrust stand consisted of a motor taken from a Cheerson CX-STARS quadcopter mounted on perpendicular strain gauges. We varied the voltage delivered to the motor in order to change the applied power, and monitored the propeller's rotational frequency using an encoder. The forces measured by the strain gauges, together with the rotational frequency information, allowed us to obtain the propeller performance coefficients (see equations presented in the Supplemental Materials). 

\begin{figure}[ht!]
    \centering
    \includegraphics[width=0.5\textwidth]{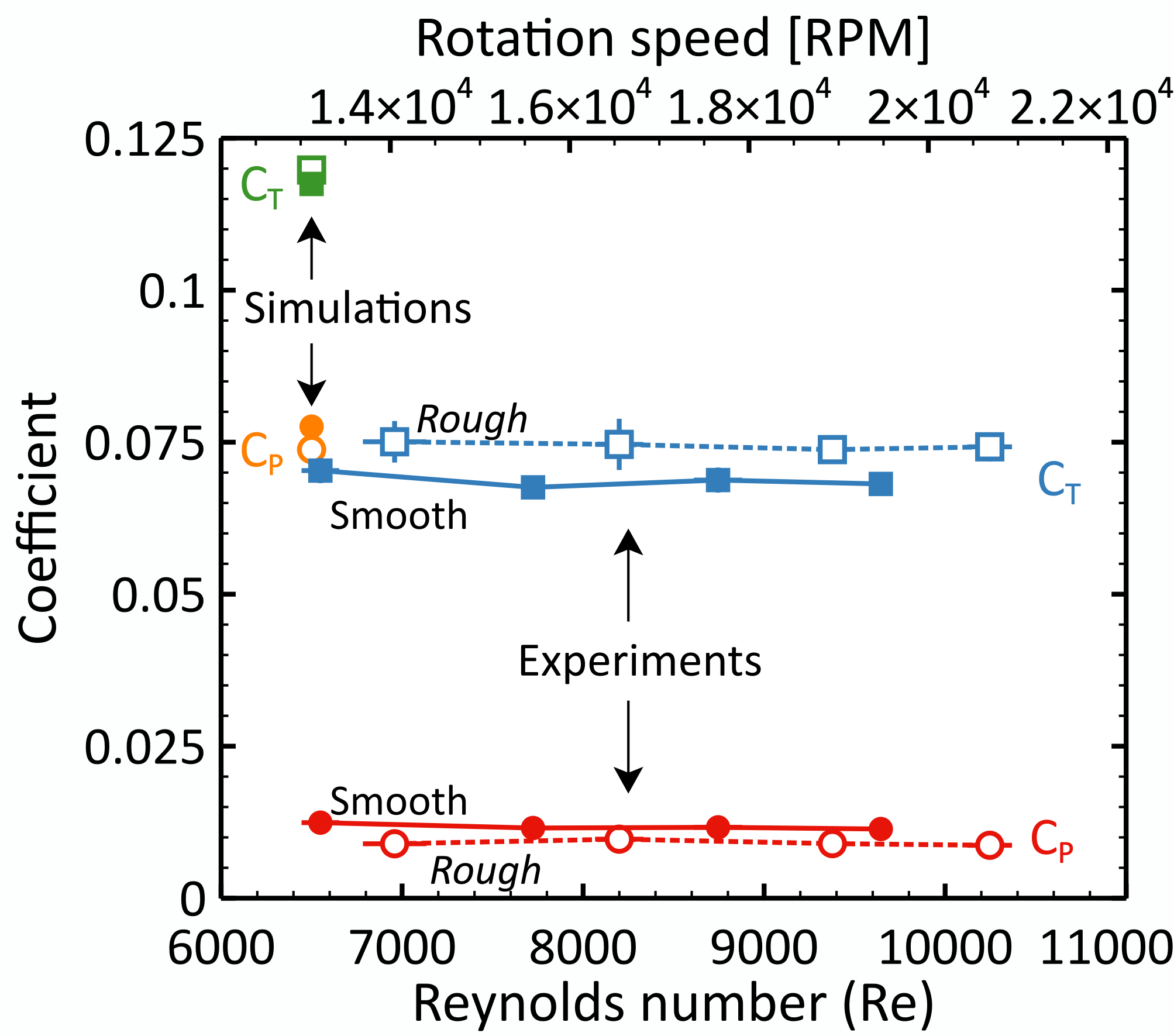}
    \caption{Comparison of experimental and numerical thrust and power coefficients for the smooth and rough propellers. {$\Bar{C}_T$ values are represented by squares, and $\Bar{C}_P$ by circles. Results for the baseline propeller have filled in symbols and are connected by continuous lines, whereas rough propeller results have open symbols and are connected by dashed lines.}}
    \label{fig:CtCpCoefPlot}
\end{figure}

The experimental results for $C_T$ (thrust coefficient) and $C_P$ (power coefficient), are summarized in Fig.~\ref{fig:CtCpCoefPlot}. We can observe that $C_T$ and $C_P$ are independent of $Re$ within the range tested. This agrees with results from Deters~\etal~\cite{deters2008static}, who measured propellers with $6<d<13$~\si{\centi\meter} in the range $10000<Re<50000$. Furthermore, the magnitudes of the $C_T$ and $C_P$ values presented in this figure are similar to those of Deters~\etal~\cite{deters2008static}{; specifically, the smallest two propellers by diameter ($6.5~\si{\centi\meter}$) had $C_T$ values between $0.06$ and $0.075$, which agree well with the present experimental values found. These facts give confidence to the experimental results.} We may also observe that both the experiments and simulations show that the rough propeller has a higher $C_T$ and a lower $C_P$ than the smooth propeller. However, the magnitudes of the coefficients differ between the experiments and simulations.  We attribute this difference to 
assumptions related to the simulations. The simulations were run in free air, with just the propeller, not including the test stand configuration. {The study by Kroo~\etal\cite{kroo2001mesicopter}, of which the baseline propeller design was based on, also found numerical thrust values for hover that over-predicted experimental results. Their discrepancies were determined to be the result of the tip of the blade operating at or near stall conditions, which would impact rotor performance. Further, the propeller thrust is not analyzed independent of any potential hub effects, which could also impact results.} Importantly, however, we emphasize that the experiments and simulations agree on the impact of the roughness; this again lends confidence to the simulation results presented in this article. Lastly, we reiterate that our intent in this work was not to optimize the propeller design but rather to demonstrate and investigate the impact of roughness at low $Re$. Future works should co-optimize both the propeller geometry and texture in order to fully leverage the roughness effect. 


\section{Conclusion}
\label{sec:conclusion}

To conclude, we carried out a numerical study to investigate the effects of surface roughness on propellers for use by micro aerial vehicles (MAVs). Two propeller blades identical in geometry and differing only in the presence of surface roughness on the top-sides of their blades were operated at a single Reynolds number $Re_c=6500$ and compared. We supplemented this numerical work by conducting thrust stand experiments on 3D-printed propellers that were identical to those studied numerically. Our numerical study determined that the addition of roughness elements to the suction side of a MAV propeller blade improved aerodynamic performance, characterized by a 2\% increase in the thrust coefficient, as well as a 5\% decrease in the power coefficient. Along with improved thrust, we found that the roughness elements induced a lower, and larger in area, pressure suction peak at the leading edge, when compared against the baseline smooth propeller. This suction peak is indicative of the leading-edge vortex, a low $Re$ flow phenomena that has been known to augment lift capabilities, and provides evidence that the roughness is enhancing these vortex structures. The experimental results corroborated the numerical results; in particular, the rough propeller showed a higher thrust coefficient and lower power coefficient than the smooth propeller.  Lastly, we note that we did not attempt an overall optimization of the roughened propeller geometry. However, the results we present provide strong motivation for future work toward this goal.  

{From this numerical study, and the accompanying experimental work, we find that the addition of surface roughness to micro-scale propellers offers the potential for improved aerodynamics at low $Re$. Therefore, we believe future MAV propeller design may benefit from inclusion of surface roughness as a design parameter. Design optimization of surface roughness is a clear, but extensive, pathway for future research. Our study investigated a single roughness style, hence future research dedicated to optimizing roughness styles would be beneficial. Another study of interest would be to examine how the location of the roughness elements impacts propeller performance, similar to that of De~Tavernier~\etal~\cite{tavernier2021controlling} for airfoils. Identifying optimal locations for roughness elements would aid in future design targeted for specific passive flow control. It is also critical to understand how surface roughness effects change with $Re$ within the low-Reynolds number regime. As shown by Lee and Jang~\cite{lee2005control}, the performance enhancements of micro-riblets on airfoils begins to depreciate as $Re$ is increased; hence, determining if these effects are not isolated to riblets is important for smart design. Recent studies have indicated that MAV noise can be decreased, at the expense of efficiency, by making slight geometric modifications to propeller blades~\cite{Jordan2020-2595, vanTreuren2019-121017, vanTreuren2020-3955}.  It may be possible to counteract the efficiency losses using roughness elements, and in so doing create an equally-efficient, but dramatically quieter propeller. To summarize, there are many other outlets for research on surface roughness for MAV propeller design.}

\section*{Funding Sources}

This work was supported by the Defense Advanced Research Projects Agency (DARPA) under Contract HR0011-19-C-0052.  

\section*{Acknowledgements}

The authors wish to thank Anna K.\ Estep, George A.\ Popov, Yuzhi Wang, Dr.\ Wujoon Cha, Prof.\ Cynthia R.\ Sung, and Prof.\ James H.\ Pikul for useful discussions and help with the propeller measurements. \textit{Any opinions, findings and conclusions or recommendations expressed in this material are those of the author(s) and do not necessarily reflect the views of the DARPA. Approved for Public Release, Distribution Unlimited.}

\nocite{*}
\bibliography{theBib}

\begin{thebibliography}{40}%
\makeatletter
\providecommand \@ifxundefined [1]{%
 \@ifx{#1\undefined}
}%
\providecommand \@ifnum [1]{%
 \ifnum #1\expandafter \@firstoftwo
 \else \expandafter \@secondoftwo
 \fi
}%
\providecommand \@ifx [1]{%
 \ifx #1\expandafter \@firstoftwo
 \else \expandafter \@secondoftwo
 \fi
}%
\providecommand \natexlab [1]{#1}%
\providecommand \enquote  [1]{``#1''}%
\providecommand \bibnamefont  [1]{#1}%
\providecommand \bibfnamefont [1]{#1}%
\providecommand \citenamefont [1]{#1}%
\providecommand \href@noop [0]{\@secondoftwo}%
\providecommand \href [0]{\begingroup \@sanitize@url \@href}%
\providecommand \@href[1]{\@@startlink{#1}\@@href}%
\providecommand \@@href[1]{\endgroup#1\@@endlink}%
\providecommand \@sanitize@url [0]{\catcode `\\12\catcode `\$12\catcode
  `\&12\catcode `\#12\catcode `\^12\catcode `\_12\catcode `\%12\relax}%
\providecommand \@@startlink[1]{}%
\providecommand \@@endlink[0]{}%
\providecommand \url  [0]{\begingroup\@sanitize@url \@url }%
\providecommand \@url [1]{\endgroup\@href {#1}{\urlprefix }}%
\providecommand \urlprefix  [0]{URL }%
\providecommand \Eprint [0]{\href }%
\providecommand \doibase [0]{http://dx.doi.org/}%
\providecommand \selectlanguage [0]{\@gobble}%
\providecommand \bibinfo  [0]{\@secondoftwo}%
\providecommand \bibfield  [0]{\@secondoftwo}%
\providecommand \translation [1]{[#1]}%
\providecommand \BibitemOpen [0]{}%
\providecommand \bibitemStop [0]{}%
\providecommand \bibitemNoStop [0]{.\EOS\space}%
\providecommand \EOS [0]{\spacefactor3000\relax}%
\providecommand \BibitemShut  [1]{\csname bibitem#1\endcsname}%
\let\auto@bib@innerbib\@empty
\bibitem [{\citenamefont {Mueller}\ and\ \citenamefont
  {DeLaurier}(2003)}]{mueller2003aerodynamics}%
  \BibitemOpen
  \bibfield  {author} {\bibinfo {author} {\bibfnamefont {T.~J.}\ \bibnamefont
  {Mueller}}\ and\ \bibinfo {author} {\bibfnamefont {J.~D.}\ \bibnamefont
  {DeLaurier}},\ }\bibfield  {title} {\enquote {\bibinfo {title} {Aerodynamics
  of small vehicles},}\ }\href {\doibase
  10.1146/annurev.fluid.35.101101.161102} {\bibfield  {journal} {\bibinfo
  {journal} {Annual Review of Fluid Mechanics}\ }\textbf {\bibinfo {volume}
  {35}},\ \bibinfo {pages} {89--111} (\bibinfo {year} {2003})}\BibitemShut
  {NoStop}%
\bibitem [{\citenamefont {Smedresman}, \citenamefont {Yeo},\ and\ \citenamefont
  {Shyy}(2011)}]{Smedresman2011Design}%
  \BibitemOpen
  \bibfield  {author} {\bibinfo {author} {\bibfnamefont {A.}~\bibnamefont
  {Smedresman}}, \bibinfo {author} {\bibfnamefont {D.}~\bibnamefont {Yeo}}, \
  and\ \bibinfo {author} {\bibfnamefont {W.}~\bibnamefont {Shyy}},\ }\bibfield
  {title} {\enquote {\bibinfo {title} {Design, fabrication, analysis, and
  testing of a micro air vehicle propeller},}\ }in\ \href {\doibase
  10.2514/6.2011-3817} {\emph {\bibinfo {booktitle} {29$^\text{th}$ AIAA
  Applied Aerodynamics Conference}}}\ (\bibinfo  {publisher} {AIAA},\ \bibinfo
  {address} {Honolulu, HI},\ \bibinfo {year} {June 2011})\ \bibinfo {note}
  {{AIAA} Paper 2011-3817}\BibitemShut {NoStop}%
\bibitem [{\citenamefont {Kroo}\ \emph {et~al.}(2001)\citenamefont {Kroo},
  \citenamefont {Prinz}, \citenamefont {Shantz}, \citenamefont {Kunz},
  \citenamefont {Fay}, \citenamefont {Cheng}, \citenamefont {Fabian},\ and\
  \citenamefont {Partridge}}]{kroo2001mesicopter}%
  \BibitemOpen
  \bibfield  {author} {\bibinfo {author} {\bibfnamefont {I.}~\bibnamefont
  {Kroo}}, \bibinfo {author} {\bibfnamefont {F.}~\bibnamefont {Prinz}},
  \bibinfo {author} {\bibfnamefont {M.}~\bibnamefont {Shantz}}, \bibinfo
  {author} {\bibfnamefont {P.}~\bibnamefont {Kunz}}, \bibinfo {author}
  {\bibfnamefont {G.}~\bibnamefont {Fay}}, \bibinfo {author} {\bibfnamefont
  {S.}~\bibnamefont {Cheng}}, \bibinfo {author} {\bibfnamefont
  {T.}~\bibnamefont {Fabian}}, \ and\ \bibinfo {author} {\bibfnamefont
  {C.}~\bibnamefont {Partridge}},\ }\bibfield  {title} {\enquote {\bibinfo
  {title} {The mesicopter: A miniature rotorcraft concept phase ii final
  report},}\ \ }(\bibinfo {address} {Stanford University, CA},\ \bibinfo {year}
  {Nov. 2001})\BibitemShut {NoStop}%
\bibitem [{\citenamefont {Deters}\ \emph {et~al.}(2018)\citenamefont {Deters},
  \citenamefont {Dantsker}, \citenamefont {Kleinke}, \citenamefont {Norman},\
  and\ \citenamefont {Selig}}]{Deters2018-4122}%
  \BibitemOpen
  \bibfield  {author} {\bibinfo {author} {\bibfnamefont {R.~W.}\ \bibnamefont
  {Deters}}, \bibinfo {author} {\bibfnamefont {O.~D.}\ \bibnamefont
  {Dantsker}}, \bibinfo {author} {\bibfnamefont {S.}~\bibnamefont {Kleinke}},
  \bibinfo {author} {\bibfnamefont {N.}~\bibnamefont {Norman}}, \ and\ \bibinfo
  {author} {\bibfnamefont {M.}~\bibnamefont {Selig}},\ }\bibfield  {title}
  {\enquote {\bibinfo {title} {Static performance results of propellers used on
  nano, micro, and mini quadrotors},}\ }in\ \href {\doibase
  10.2514/6.2018-4122} {\emph {\bibinfo {booktitle} {2018 Applied Aerodynamics
  Conference}}}\ (\bibinfo {address} {Atlanta, GA},\ \bibinfo {year} {June
  2018})\ \bibinfo {note} {{AIAA} Paper 2018-4122}\BibitemShut {NoStop}%
\bibitem [{\citenamefont {Deters}\ and\ \citenamefont
  {Selig}(2008)}]{deters2008static}%
  \BibitemOpen
  \bibfield  {author} {\bibinfo {author} {\bibfnamefont {R.~W.}\ \bibnamefont
  {Deters}}\ and\ \bibinfo {author} {\bibfnamefont {M.~S.}\ \bibnamefont
  {Selig}},\ }\bibfield  {title} {\enquote {\bibinfo {title} {Static testing of
  micro propellers},}\ }in\ \href {\doibase 10.2514/6.2008-6246} {\emph
  {\bibinfo {booktitle} {26$^{\text{th}}$ AIAA Applied Aerodynamics
  Conference}}}\ (\bibinfo {address} {Honolulu, HI},\ \bibinfo {year} {Aug.
  2008})\ \bibinfo {note} {{AIAA} Paper 2008-6246}\BibitemShut {NoStop}%
\bibitem [{\citenamefont {Deters}, \citenamefont {Ananda},\ and\ \citenamefont
  {Selig}(2014)}]{deters2014reynolds}%
  \BibitemOpen
  \bibfield  {author} {\bibinfo {author} {\bibfnamefont {R.~W.}\ \bibnamefont
  {Deters}}, \bibinfo {author} {\bibfnamefont {G.~K.}\ \bibnamefont {Ananda}},
  \ and\ \bibinfo {author} {\bibfnamefont {M.~S.}\ \bibnamefont {Selig}},\
  }\bibfield  {title} {\enquote {\bibinfo {title} {Reynolds number effects on
  the performance of small-scale propellers},}\ }in\ \href {\doibase
  10.2514/6.2014-2151} {\emph {\bibinfo {booktitle} {32$^{\text{nd}}$ AIAA
  Applied Aerodynamics Conference}}}\ (\bibinfo {address} {Atlanta, GA},\
  \bibinfo {year} {Jun. 2014})\ \bibinfo {note} {{AIAA} Paper
  2014-2151}\BibitemShut {NoStop}%
\bibitem [{\citenamefont {Selig}, \citenamefont {Deters},\ and\ \citenamefont
  {Williamson}(2011)}]{selig2011wind}%
  \BibitemOpen
  \bibfield  {author} {\bibinfo {author} {\bibfnamefont {M.~S.}\ \bibnamefont
  {Selig}}, \bibinfo {author} {\bibfnamefont {R.~W.}\ \bibnamefont {Deters}}, \
  and\ \bibinfo {author} {\bibfnamefont {G.~A.}\ \bibnamefont {Williamson}},\
  }\bibfield  {title} {\enquote {\bibinfo {title} {Wind tunnel testing airfoils
  at low reynolds numbers},}\ }in\ \href {\doibase 10.2514/6.2011-875} {\emph
  {\bibinfo {booktitle} {49$^\text{th}$ AIAA Aerospace Sciences Meeting}}}\
  (\bibinfo {address} {Orlando, FL},\ \bibinfo {year} {Jan. 2011})\ \bibinfo
  {note} {{AIAA} Paper 2011-875}\BibitemShut {NoStop}%
\bibitem [{\citenamefont {Devey}\ \emph {et~al.}(2020)\citenamefont {Devey},
  \citenamefont {Lang}, \citenamefont {Hubner}, \citenamefont {Morris},\ and\
  \citenamefont {Habegger}}]{devey2020experimental}%
  \BibitemOpen
  \bibfield  {author} {\bibinfo {author} {\bibfnamefont {S.~P.}\ \bibnamefont
  {Devey}}, \bibinfo {author} {\bibfnamefont {A.~W.}\ \bibnamefont {Lang}},
  \bibinfo {author} {\bibfnamefont {J.~P.}\ \bibnamefont {Hubner}}, \bibinfo
  {author} {\bibfnamefont {J.~A.}\ \bibnamefont {Morris}}, \ and\ \bibinfo
  {author} {\bibfnamefont {M.~L.}\ \bibnamefont {Habegger}},\ }\bibfield
  {title} {\enquote {\bibinfo {title} {Experimental analysis of passive
  bristling in air to enable mako-shark-inspired separation control},}\ }in\
  \href {\doibase 10.2514/6.2020-2768} {\emph {\bibinfo {booktitle} {AIAA
  Aviation 2020 Forum}}}\ (\bibinfo {address} {Virtual},\ \bibinfo {year} {June
  2020})\ \bibinfo {note} {{AIAA} Paper 2020-2768}\BibitemShut {NoStop}%
\bibitem [{\citenamefont {Asghar}\ \emph {et~al.}(2020)\citenamefont {Asghar},
  \citenamefont {Perez}, \citenamefont {Jansen},\ and\ \citenamefont
  {Allan}}]{asghar2020application}%
  \BibitemOpen
  \bibfield  {author} {\bibinfo {author} {\bibfnamefont {A.}~\bibnamefont
  {Asghar}}, \bibinfo {author} {\bibfnamefont {R.~E.}\ \bibnamefont {Perez}},
  \bibinfo {author} {\bibfnamefont {P.~W.}\ \bibnamefont {Jansen}}, \ and\
  \bibinfo {author} {\bibfnamefont {W.~D.~E.}\ \bibnamefont {Allan}},\
  }\bibfield  {title} {\enquote {\bibinfo {title} {Application of leading-edge
  tubercles to enhance propeller performance},}\ }\href {\doibase
  10.2514/1.J058740} {\bibfield  {journal} {\bibinfo  {journal} {AIAA Journal}\
  }\textbf {\bibinfo {volume} {58}},\ \bibinfo {pages} {4659--4671} (\bibinfo
  {year} {2020})}\BibitemShut {NoStop}%
\bibitem [{\citenamefont {Butt}\ and\ \citenamefont
  {Talha}(2019)}]{butt2019numerical}%
  \BibitemOpen
  \bibfield  {author} {\bibinfo {author} {\bibfnamefont {F.~R.}\ \bibnamefont
  {Butt}}\ and\ \bibinfo {author} {\bibfnamefont {T.}~\bibnamefont {Talha}},\
  }\bibfield  {title} {\enquote {\bibinfo {title} {Numerical investigation of
  the effect of leading-edge tubercles on propeller performance},}\ }\href
  {\doibase 10.2514/1.C034845} {\bibfield  {journal} {\bibinfo  {journal}
  {Journal of Aircraft}\ }\textbf {\bibinfo {volume} {56}},\ \bibinfo {pages}
  {1014--1028} (\bibinfo {year} {2019})}\BibitemShut {NoStop}%
\bibitem [{\citenamefont {Zhou}\ and\ \citenamefont
  {Wang}(2012)}]{zhou2012effects}%
  \BibitemOpen
  \bibfield  {author} {\bibinfo {author} {\bibfnamefont {Y.}~\bibnamefont
  {Zhou}}\ and\ \bibinfo {author} {\bibfnamefont {Z.~J.}\ \bibnamefont
  {Wang}},\ }\bibfield  {title} {\enquote {\bibinfo {title} {Effects of surface
  roughness on separated and transitional flows over a wing},}\ }\href
  {\doibase 10.2514/1.J051237} {\bibfield  {journal} {\bibinfo  {journal} {AIAA
  Journal}\ }\textbf {\bibinfo {volume} {50}},\ \bibinfo {pages} {593--609}
  (\bibinfo {year} {2012})}\BibitemShut {NoStop}%
\bibitem [{\citenamefont {Lee}\ and\ \citenamefont
  {Jang}(2005)}]{lee2005control}%
  \BibitemOpen
  \bibfield  {author} {\bibinfo {author} {\bibfnamefont {S.~J.}\ \bibnamefont
  {Lee}}\ and\ \bibinfo {author} {\bibfnamefont {Y.~G.}\ \bibnamefont {Jang}},\
  }\bibfield  {title} {\enquote {\bibinfo {title} {Control of flow around a
  naca 0012 airfoil with a micro-riblet film},}\ }\href {\doibase
  10.1016/j.jfluidstructs.2005.03.003} {\bibfield  {journal} {\bibinfo
  {journal} {Journal of Fluids and Structures}\ }\textbf {\bibinfo {volume}
  {20}},\ \bibinfo {pages} {659--672} (\bibinfo {year} {2005})}\BibitemShut
  {NoStop}%
\bibitem [{\citenamefont {Mukherjee}\ \emph {et~al.}(2021)\citenamefont
  {Mukherjee}, \citenamefont {Pawar}, \citenamefont {Ranjan},\ and\
  \citenamefont {Saha}}]{mukherjee2021corrugation}%
  \BibitemOpen
  \bibfield  {author} {\bibinfo {author} {\bibfnamefont {P.}~\bibnamefont
  {Mukherjee}}, \bibinfo {author} {\bibfnamefont {A.~A.}\ \bibnamefont
  {Pawar}}, \bibinfo {author} {\bibfnamefont {K.~S.}\ \bibnamefont {Ranjan}}, \
  and\ \bibinfo {author} {\bibfnamefont {S.}~\bibnamefont {Saha}},\ }\bibfield
  {title} {\enquote {\bibinfo {title} {Corrugation assisted enhancement of
  aerodynamic characteristics of delta wing for micro aerial vehicle},}\ }\href
  {\doibase 10.2514/1.C035849} {\bibfield  {journal} {\bibinfo  {journal}
  {Journal of Aircraft}\ }\textbf {\bibinfo {volume} {58}} (\bibinfo {year}
  {2021}),\ 10.2514/1.C035849}\BibitemShut {NoStop}%
\bibitem [{\citenamefont {Heine}\ \emph {et~al.}(2013)\citenamefont {Heine},
  \citenamefont {Mulleners}, \citenamefont {Joubert},\ and\ \citenamefont
  {Raffel}}]{heine2013dynamic}%
  \BibitemOpen
  \bibfield  {author} {\bibinfo {author} {\bibfnamefont {B.}~\bibnamefont
  {Heine}}, \bibinfo {author} {\bibfnamefont {K.}~\bibnamefont {Mulleners}},
  \bibinfo {author} {\bibfnamefont {G.}~\bibnamefont {Joubert}}, \ and\
  \bibinfo {author} {\bibfnamefont {M.}~\bibnamefont {Raffel}},\ }\bibfield
  {title} {\enquote {\bibinfo {title} {Dynamic stall control by passive
  disturbance generators},}\ }\href {\doibase 10.2514/1.J051525} {\bibfield
  {journal} {\bibinfo  {journal} {AIAA Journal}\ }\textbf {\bibinfo {volume}
  {51}},\ \bibinfo {pages} {2086--2097} (\bibinfo {year} {2013})}\BibitemShut
  {NoStop}%
\bibitem [{\citenamefont {De~Tavernier}\ \emph {et~al.}(2021)\citenamefont
  {De~Tavernier}, \citenamefont {Ferreira}, \citenamefont {Vir\'e},
  \citenamefont {LeBlanc},\ and\ \citenamefont
  {Bernardy}}]{tavernier2021controlling}%
  \BibitemOpen
  \bibfield  {author} {\bibinfo {author} {\bibfnamefont {D.}~\bibnamefont
  {De~Tavernier}}, \bibinfo {author} {\bibfnamefont {C.}~\bibnamefont
  {Ferreira}}, \bibinfo {author} {\bibfnamefont {A.}~\bibnamefont {Vir\'e}},
  \bibinfo {author} {\bibfnamefont {B.}~\bibnamefont {LeBlanc}}, \ and\
  \bibinfo {author} {\bibfnamefont {S.}~\bibnamefont {Bernardy}},\ }\bibfield
  {title} {\enquote {\bibinfo {title} {Controlling dynamic stall using vortex
  generators on a wind turbine airfoil},}\ }\href {\doibase
  10.1016/j.renene.2021.03.019} {\bibfield  {journal} {\bibinfo  {journal}
  {Renewable Energy}\ }\textbf {\bibinfo {volume} {172}},\ \bibinfo {pages}
  {1194--1211} (\bibinfo {year} {2021})}\BibitemShut {NoStop}%
\bibitem [{\citenamefont {Lentink}\ \emph {et~al.}(2009)\citenamefont
  {Lentink}, \citenamefont {Dickson}, \citenamefont {Van~Leeuwen},\ and\
  \citenamefont {Dickinson}}]{lentink2009leading}%
  \BibitemOpen
  \bibfield  {author} {\bibinfo {author} {\bibfnamefont {D.}~\bibnamefont
  {Lentink}}, \bibinfo {author} {\bibfnamefont {W.~B.}\ \bibnamefont
  {Dickson}}, \bibinfo {author} {\bibfnamefont {J.~L.}\ \bibnamefont
  {Van~Leeuwen}}, \ and\ \bibinfo {author} {\bibfnamefont {M.~H.}\ \bibnamefont
  {Dickinson}},\ }\bibfield  {title} {\enquote {\bibinfo {title} {Leading-edge
  vortices elevate lift of autorotating plant seeds},}\ }\href {\doibase
  10.1126/science.1174196} {\bibfield  {journal} {\bibinfo  {journal}
  {Science}\ }\textbf {\bibinfo {volume} {325}},\ \bibinfo {pages} {1438--1440}
  (\bibinfo {year} {2009})}\BibitemShut {NoStop}%
\bibitem [{\citenamefont {Pitt~Ford}\ and\ \citenamefont
  {Babinksy}(2013)}]{ford2013lift}%
  \BibitemOpen
  \bibfield  {author} {\bibinfo {author} {\bibfnamefont {C.~W.}\ \bibnamefont
  {Pitt~Ford}}\ and\ \bibinfo {author} {\bibfnamefont {H.}~\bibnamefont
  {Babinksy}},\ }\bibfield  {title} {\enquote {\bibinfo {title} {Lift and the
  leading-edge vortex},}\ }\href {\doibase 10.1017/jfm.2013.28} {\bibfield
  {journal} {\bibinfo  {journal} {Journal of Fluid Mechanics}\ }\textbf
  {\bibinfo {volume} {720}},\ \bibinfo {pages} {280--313} (\bibinfo {year}
  {2013})}\BibitemShut {NoStop}%
\bibitem [{\citenamefont {Lee}\ and\ \citenamefont
  {Choi}(2017)}]{lee2017flight}%
  \BibitemOpen
  \bibfield  {author} {\bibinfo {author} {\bibfnamefont {I.}~\bibnamefont
  {Lee}}\ and\ \bibinfo {author} {\bibfnamefont {H.}~\bibnamefont {Choi}},\
  }\bibfield  {title} {\enquote {\bibinfo {title} {Flight of a falling maple
  seed},}\ }\href {\doibase 10.1103/PhysRevFluids.2.090511} {\bibfield
  {journal} {\bibinfo  {journal} {Physical Review Fluids}\ }\textbf {\bibinfo
  {volume} {2}},\ \bibinfo {pages} {090511} (\bibinfo {year}
  {2017})}\BibitemShut {NoStop}%
\bibitem [{\citenamefont {Jardin}, \citenamefont {Choi},\ and\ \citenamefont
  {Colonius}(2021)}]{jardin2021empirical}%
  \BibitemOpen
  \bibfield  {author} {\bibinfo {author} {\bibfnamefont {T.}~\bibnamefont
  {Jardin}}, \bibinfo {author} {\bibfnamefont {J.}~\bibnamefont {Choi}}, \ and\
  \bibinfo {author} {\bibfnamefont {T.}~\bibnamefont {Colonius}},\ }\bibfield
  {title} {\enquote {\bibinfo {title} {An empirical correlation between lift
  and the properties of leading-edge vortices},}\ }\href {\doibase
  10.1007/s00162-021-00567-x} {\bibfield  {journal} {\bibinfo  {journal}
  {Theoretical and Computational Fluid Dynamics}\ }\textbf {\bibinfo {volume}
  {35}},\ \bibinfo {pages} {437--448} (\bibinfo {year} {2021})}\BibitemShut
  {NoStop}%
\bibitem [{\citenamefont {Jardin}(2017)}]{jardin2017Coriolis}%
  \BibitemOpen
  \bibfield  {author} {\bibinfo {author} {\bibfnamefont {T.}~\bibnamefont
  {Jardin}},\ }\bibfield  {title} {\enquote {\bibinfo {title} {Coriolis effect
  and the attachment of the leading edge vortex},}\ }\href {\doibase
  10.1017/jfm.2017.222} {\bibfield  {journal} {\bibinfo  {journal} {Journal of
  Fluid Mechanics}\ }\textbf {\bibinfo {volume} {820}},\ \bibinfo {pages}
  {312--340} (\bibinfo {year} {2017})}\BibitemShut {NoStop}%
\bibitem [{\citenamefont {Ukken}\ and\ \citenamefont
  {Sivapragasam}(2019)}]{Ukken2019-130}%
  \BibitemOpen
  \bibfield  {author} {\bibinfo {author} {\bibfnamefont {M.~G.}\ \bibnamefont
  {Ukken}}\ and\ \bibinfo {author} {\bibfnamefont {M.}~\bibnamefont
  {Sivapragasam}},\ }\bibfield  {title} {\enquote {\bibinfo {title}
  {Aerodynamic shape optimization of airfoils at ultra-low reynolds numbers},}\
  }\href {\doibase 10.1007/s12046-019-1115-z} {\bibfield  {journal} {\bibinfo
  {journal} {S{\=a}dhan{\=a}}\ }\textbf {\bibinfo {volume} {44}} (\bibinfo
  {year} {2019}),\ 10.1007/s12046-019-1115-z}\BibitemShut {NoStop}%
\bibitem [{\citenamefont {Jim\'enez}(2004)}]{jimenez2004turbulent}%
  \BibitemOpen
  \bibfield  {author} {\bibinfo {author} {\bibfnamefont {J.}~\bibnamefont
  {Jim\'enez}},\ }\bibfield  {title} {\enquote {\bibinfo {title} {Turbulent
  flows over rough walls},}\ }\href {\doibase
  10.1146/annurev.fluid.36.050802.122103} {\bibfield  {journal} {\bibinfo
  {journal} {Annual Review of Fluid Mechanics}\ }\textbf {\bibinfo {volume}
  {36}},\ \bibinfo {pages} {173--196} (\bibinfo {year} {2004})}\BibitemShut
  {NoStop}%
\bibitem [{\citenamefont {Goc}, \citenamefont {Moin},\ and\ \citenamefont
  {Bose}(2020)}]{goc20201wmles}%
  \BibitemOpen
  \bibfield  {author} {\bibinfo {author} {\bibfnamefont {K.~A.}\ \bibnamefont
  {Goc}}, \bibinfo {author} {\bibfnamefont {P.}~\bibnamefont {Moin}}, \ and\
  \bibinfo {author} {\bibfnamefont {S.~T.}\ \bibnamefont {Bose}},\ }\bibfield
  {title} {\enquote {\bibinfo {title} {Wall-modeled large eddy simulation of an
  aircraft in landing configuration},}\ }in\ \href {\doibase
  10.2514/6.2020-3002} {\emph {\bibinfo {booktitle} {AIAA Aviation 2020
  Forum}}}\ (\bibinfo {address} {Virtual},\ \bibinfo {year} {June 2020})\
  \bibinfo {note} {{AIAA} Paper 2020-3002}\BibitemShut {NoStop}%
\bibitem [{\citenamefont {Br\`es}\ \emph {et~al.}(2017)\citenamefont {Br\`es},
  \citenamefont {Ham}, \citenamefont {Nichols},\ and\ \citenamefont
  {Lele}}]{bres2017unstructured}%
  \BibitemOpen
  \bibfield  {author} {\bibinfo {author} {\bibfnamefont {G.~A.}\ \bibnamefont
  {Br\`es}}, \bibinfo {author} {\bibfnamefont {F.~E.}\ \bibnamefont {Ham}},
  \bibinfo {author} {\bibfnamefont {J.~W.}\ \bibnamefont {Nichols}}, \ and\
  \bibinfo {author} {\bibfnamefont {S.~K.}\ \bibnamefont {Lele}},\ }\bibfield
  {title} {\enquote {\bibinfo {title} {Unstructured large-eddy simulations of
  supersonic jets},}\ }\href {\doibase 10.2514/1.J055084} {\bibfield  {journal}
  {\bibinfo  {journal} {AIAA Journal}\ }\textbf {\bibinfo {volume} {55}},\
  \bibinfo {pages} {1164--1184} (\bibinfo {year} {2017})}\BibitemShut {NoStop}%
\bibitem [{\citenamefont {Lozano-Dur\'an}, \citenamefont {Bose},\ and\
  \citenamefont {Moin}(2022)}]{lozanoduran2022performance}%
  \BibitemOpen
  \bibfield  {author} {\bibinfo {author} {\bibfnamefont {A.}~\bibnamefont
  {Lozano-Dur\'an}}, \bibinfo {author} {\bibfnamefont {S.~T.}\ \bibnamefont
  {Bose}}, \ and\ \bibinfo {author} {\bibfnamefont {P.}~\bibnamefont {Moin}},\
  }\bibfield  {title} {\enquote {\bibinfo {title} {Performance of wall-modeled
  les with boundary-layer-conforming grids for external aerodynamics},}\ }\href
  {\doibase 10.2514/1.J061041} {\bibfield  {journal} {\bibinfo  {journal} {AIAA
  Journal}\ }\textbf {\bibinfo {volume} {60}},\ \bibinfo {pages} {747--766}
  (\bibinfo {year} {2022})}\BibitemShut {NoStop}%
\bibitem [{\citenamefont {Powers}\ and\ \citenamefont
  {Gilbert}(2023)}]{powers2023unsteady}%
  \BibitemOpen
  \bibfield  {author} {\bibinfo {author} {\bibfnamefont {R.}~\bibnamefont
  {Powers}}\ and\ \bibinfo {author} {\bibfnamefont {J.}~\bibnamefont
  {Gilbert}},\ }\bibfield  {title} {\enquote {\bibinfo {title} {Unsteady
  simulation of a single stage high speed compressor with advanced casing
  treatments},}\ }in\ \href {\doibase 10.2514/6.2023-2652} {\emph {\bibinfo
  {booktitle} {AIAA SciTech Forum}}}\ (\bibinfo {address} {National Harbor, MD
  and Online},\ \bibinfo {year} {Jan. 2023})\ \bibinfo {note} {{AIAA} Paper
  2023-2652}\BibitemShut {NoStop}%
\bibitem [{\citenamefont {Jain}\ \emph {et~al.}(2020)\citenamefont {Jain},
  \citenamefont {Bravo}, \citenamefont {Kim}, \citenamefont {Murugan},
  \citenamefont {Ghoshal}, \citenamefont {Ham},\ and\ \citenamefont
  {Flatau}}]{jain2020massively}%
  \BibitemOpen
  \bibfield  {author} {\bibinfo {author} {\bibfnamefont {N.}~\bibnamefont
  {Jain}}, \bibinfo {author} {\bibfnamefont {L.}~\bibnamefont {Bravo}},
  \bibinfo {author} {\bibfnamefont {D.}~\bibnamefont {Kim}}, \bibinfo {author}
  {\bibfnamefont {M.}~\bibnamefont {Murugan}}, \bibinfo {author} {\bibfnamefont
  {A.}~\bibnamefont {Ghoshal}}, \bibinfo {author} {\bibfnamefont
  {F.}~\bibnamefont {Ham}}, \ and\ \bibinfo {author} {\bibfnamefont
  {A.}~\bibnamefont {Flatau}},\ }\bibfield  {title} {\enquote {\bibinfo {title}
  {Massively parallel large eddy simulation of rotating turbomachinery for
  variable speed gas turbine engine operation},}\ }\href {\doibase
  10.3390/en13030703} {\bibfield  {journal} {\bibinfo  {journal} {Energies}\
  }\textbf {\bibinfo {volume} {13}},\ \bibinfo {pages} {703} (\bibinfo {year}
  {2020})}\BibitemShut {NoStop}%
\bibitem [{\citenamefont {Kumar}\ and\ \citenamefont
  {Mahesh}(2017)}]{kumar2017large}%
  \BibitemOpen
  \bibfield  {author} {\bibinfo {author} {\bibfnamefont {P.}~\bibnamefont
  {Kumar}}\ and\ \bibinfo {author} {\bibfnamefont {K.}~\bibnamefont {Mahesh}},\
  }\bibfield  {title} {\enquote {\bibinfo {title} {Large eddy simulation of
  propeller wake instabilities},}\ }\href {\doibase 10.1017/jfm.2017.20}
  {\bibfield  {journal} {\bibinfo  {journal} {Journal of Fluid Mechanics}\
  }\textbf {\bibinfo {volume} {814}},\ \bibinfo {pages} {361--396} (\bibinfo
  {year} {2017})}\BibitemShut {NoStop}%
\bibitem [{\citenamefont {Liu}\ \emph {et~al.}(2020)\citenamefont {Liu},
  \citenamefont {Yuan}, \citenamefont {Qiu}, \citenamefont {Feng},
  \citenamefont {Xie}, \citenamefont {Leng},\ and\ \citenamefont
  {Tian}}]{liu2020brief}%
  \BibitemOpen
  \bibfield  {author} {\bibinfo {author} {\bibfnamefont {G.}~\bibnamefont
  {Liu}}, \bibinfo {author} {\bibfnamefont {Z.}~\bibnamefont {Yuan}}, \bibinfo
  {author} {\bibfnamefont {Z.}~\bibnamefont {Qiu}}, \bibinfo {author}
  {\bibfnamefont {S.}~\bibnamefont {Feng}}, \bibinfo {author} {\bibfnamefont
  {Y.}~\bibnamefont {Xie}}, \bibinfo {author} {\bibfnamefont {D.}~\bibnamefont
  {Leng}}, \ and\ \bibinfo {author} {\bibfnamefont {X.}~\bibnamefont {Tian}},\
  }\bibfield  {title} {\enquote {\bibinfo {title} {A brief review of
  bio-inspired surface technology and application toward underwater drag
  reduction},}\ }\href {\doibase 10.1016/j.oceaneng.2020.106962} {\bibfield
  {journal} {\bibinfo  {journal} {Ocean Engineering}\ }\textbf {\bibinfo
  {volume} {199}},\ \bibinfo {pages} {106962--1--106962--8} (\bibinfo {year}
  {2020})}\BibitemShut {NoStop}%
\bibitem [{\citenamefont {Yue}\ \emph {et~al.}(2021)\citenamefont {Yue},
  \citenamefont {C}, \citenamefont {Kim}, \citenamefont {Kohlmeyer},
  \citenamefont {Patra}, \citenamefont {Grzyb}, \citenamefont {Padmanabha},
  \citenamefont {Wang}, \citenamefont {Jiang}, \citenamefont {Sun},
  \citenamefont {Kiggins}, \citenamefont {Ates}, \citenamefont {Sing},
  \citenamefont {Beale}, \citenamefont {Daroux}, \citenamefont {Blake},
  \citenamefont {Cook}, \citenamefont {Braun},\ and\ \citenamefont
  {Pikul}}]{yue2021nearly}%
  \BibitemOpen
  \bibfield  {author} {\bibinfo {author} {\bibfnamefont {X.}~\bibnamefont
  {Yue}}, \bibinfo {author} {\bibfnamefont {J.~A.}\ \bibnamefont {C}}, \bibinfo
  {author} {\bibfnamefont {S.}~\bibnamefont {Kim}}, \bibinfo {author}
  {\bibfnamefont {R.~R.}\ \bibnamefont {Kohlmeyer}}, \bibinfo {author}
  {\bibfnamefont {A.}~\bibnamefont {Patra}}, \bibinfo {author} {\bibfnamefont
  {J.}~\bibnamefont {Grzyb}}, \bibinfo {author} {\bibfnamefont
  {A.}~\bibnamefont {Padmanabha}}, \bibinfo {author} {\bibfnamefont
  {M.}~\bibnamefont {Wang}}, \bibinfo {author} {\bibfnamefont {Z.}~\bibnamefont
  {Jiang}}, \bibinfo {author} {\bibfnamefont {P.}~\bibnamefont {Sun}}, \bibinfo
  {author} {\bibfnamefont {C.~T.}\ \bibnamefont {Kiggins}}, \bibinfo {author}
  {\bibfnamefont {M.~N.}\ \bibnamefont {Ates}}, \bibinfo {author}
  {\bibfnamefont {S.~V.}\ \bibnamefont {Sing}}, \bibinfo {author}
  {\bibfnamefont {E.~M.}\ \bibnamefont {Beale}}, \bibinfo {author}
  {\bibfnamefont {M.}~\bibnamefont {Daroux}}, \bibinfo {author} {\bibfnamefont
  {A.~J.}\ \bibnamefont {Blake}}, \bibinfo {author} {\bibfnamefont {J.~B.}\
  \bibnamefont {Cook}}, \bibinfo {author} {\bibfnamefont {P.~V.}\ \bibnamefont
  {Braun}}, \ and\ \bibinfo {author} {\bibfnamefont {J.~H.}\ \bibnamefont
  {Pikul}},\ }\bibfield  {title} {\enquote {\bibinfo {title} {A nearly
  packaging-free design paradigm for light, powerful, and energy-dense primary
  microbatteries},}\ }\href {\doibase 10.1002/adma.202101760} {\bibfield
  {journal} {\bibinfo  {journal} {Advanced Materials}\ }\textbf {\bibinfo
  {volume} {33}},\ \bibinfo {pages} {2101760--1--2101760--9} (\bibinfo {year}
  {2021})}\BibitemShut {NoStop}%
\bibitem [{\citenamefont {Choi}, \citenamefont {Jeon},\ and\ \citenamefont
  {Choi}(2006)}]{choi2006mechanism}%
  \BibitemOpen
  \bibfield  {author} {\bibinfo {author} {\bibfnamefont {J.}~\bibnamefont
  {Choi}}, \bibinfo {author} {\bibfnamefont {W.~P.}\ \bibnamefont {Jeon}}, \
  and\ \bibinfo {author} {\bibfnamefont {H.}~\bibnamefont {Choi}},\ }\bibfield
  {title} {\enquote {\bibinfo {title} {Mechanism of drag reduction by dimples
  on a sphere},}\ }\href {\doibase 10.1063/1.2191848} {\bibfield  {journal}
  {\bibinfo  {journal} {Physics of Fluids}\ }\textbf {\bibinfo {volume} {18}}
  (\bibinfo {year} {2006}),\ 10.1063/1.2191848},\ \bibinfo {note} {{Paper}
  041702}\BibitemShut {NoStop}%
\bibitem [{\citenamefont {Arena}\ and\ \citenamefont
  {J}(1980)}]{arena1980laminar}%
  \BibitemOpen
  \bibfield  {author} {\bibinfo {author} {\bibfnamefont {A.~V.}\ \bibnamefont
  {Arena}}\ and\ \bibinfo {author} {\bibfnamefont {M.~T.}\ \bibnamefont {J}},\
  }\bibfield  {title} {\enquote {\bibinfo {title} {Laminar separation,
  transition, and turbulent reattachment near the leading edge of airfoils},}\
  }\href {\doibase 10.2514/3.50815} {\bibfield  {journal} {\bibinfo  {journal}
  {AIAA Journal}\ }\textbf {\bibinfo {volume} {18}},\ \bibinfo {pages}
  {747--753} (\bibinfo {year} {1980})}\BibitemShut {NoStop}%
\bibitem [{\citenamefont {Ferreira}\ and\ \citenamefont
  {Ganapathisubramani}(2020)}]{ferreira2020piv}%
  \BibitemOpen
  \bibfield  {author} {\bibinfo {author} {\bibfnamefont {M.~A.}\ \bibnamefont
  {Ferreira}}\ and\ \bibinfo {author} {\bibfnamefont {B.}~\bibnamefont
  {Ganapathisubramani}},\ }\bibfield  {title} {\enquote {\bibinfo {title}
  {Piv-based pressure estimation in the canopy of urban-like roughness},}\
  }\href {\doibase 10.1007/s00348-020-2904-1} {\bibfield  {journal} {\bibinfo
  {journal} {Experiments in Fluids}\ }\textbf {\bibinfo {volume} {61}}
  (\bibinfo {year} {2020}),\ 10.1007/s00348-020-2904-1}\BibitemShut {NoStop}%
\bibitem [{\citenamefont {Hunt}, \citenamefont {Wray},\ and\ \citenamefont
  {Moin}(1988)}]{hunt1988eddies}%
  \BibitemOpen
  \bibfield  {author} {\bibinfo {author} {\bibfnamefont {J.~C.~R.}\
  \bibnamefont {Hunt}}, \bibinfo {author} {\bibfnamefont {A.~A.}\ \bibnamefont
  {Wray}}, \ and\ \bibinfo {author} {\bibfnamefont {P.}~\bibnamefont {Moin}},\
  }\bibfield  {title} {\enquote {\bibinfo {title} {Eddies, streams, and
  convergence zones in turbulent flows},}\ }in\ \href@noop {} {\emph {\bibinfo
  {booktitle} {CTR Proceedings of the Summer Program 1988}}}\ (\bibinfo
  {address} {Stanford University, CA},\ \bibinfo {year} {1988})\BibitemShut
  {NoStop}%
\bibitem [{\citenamefont {Nabawy}\ and\ \citenamefont
  {Crowther}(2014)}]{nabawy2014quasi}%
  \BibitemOpen
  \bibfield  {author} {\bibinfo {author} {\bibfnamefont {M.~R.~A.}\
  \bibnamefont {Nabawy}}\ and\ \bibinfo {author} {\bibfnamefont {W.~J.}\
  \bibnamefont {Crowther}},\ }\bibfield  {title} {\enquote {\bibinfo {title}
  {On the quasi-steady aerodynamics of normal hovering flight part ii: Model
  implementation and evaluation},}\ }\href {\doibase 10.1098/rsif.2013.1197}
  {\bibfield  {journal} {\bibinfo  {journal} {Journal of the Royal Society
  Interface}\ }\textbf {\bibinfo {volume} {11}} (\bibinfo {year} {2014}),\
  10.1098/rsif.2013.1197}\BibitemShut {NoStop}%
\bibitem [{\citenamefont {Jordan}, \citenamefont {Narsipur},\ and\
  \citenamefont {Deters}(2020)}]{Jordan2020-2595}%
  \BibitemOpen
  \bibfield  {author} {\bibinfo {author} {\bibfnamefont {W.~A.}\ \bibnamefont
  {Jordan}}, \bibinfo {author} {\bibfnamefont {S.}~\bibnamefont {Narsipur}}, \
  and\ \bibinfo {author} {\bibfnamefont {R.}~\bibnamefont {Deters}},\
  }\bibfield  {title} {\enquote {\bibinfo {title} {Aerodynamic and aeroacoustic
  performance of small uav propellers in static conditions},}\ }in\ \href
  {\doibase 10.2514/6.2020-2595} {\emph {\bibinfo {booktitle} {AIAA AVIATION
  2020 Forum}}}\ (\bibinfo {address} {Virtual},\ \bibinfo {year} {June 2020})\
  \bibinfo {note} {{AIAA} Paper 2020-2595}\BibitemShut {NoStop}%
\bibitem [{\citenamefont {Kruyt}\ \emph {et~al.}(2014)\citenamefont {Kruyt},
  \citenamefont {Quicaz\'{a}n-{R}ubio}, \citenamefont {{van Heijst}},
  \citenamefont {Altshuler},\ and\ \citenamefont
  {Lentink}}]{Kruyt2014-20140585}%
  \BibitemOpen
  \bibfield  {author} {\bibinfo {author} {\bibfnamefont {J.~W.}\ \bibnamefont
  {Kruyt}}, \bibinfo {author} {\bibfnamefont {E.~M.}\ \bibnamefont
  {Quicaz\'{a}n-{R}ubio}}, \bibinfo {author} {\bibfnamefont {G.~F.}\
  \bibnamefont {{van Heijst}}}, \bibinfo {author} {\bibfnamefont {D.~L.}\
  \bibnamefont {Altshuler}}, \ and\ \bibinfo {author} {\bibfnamefont
  {D.}~\bibnamefont {Lentink}},\ }\bibfield  {title} {\enquote {\bibinfo
  {title} {Hummingbird wing efficacy depends on aspect ratio and compares with
  helicopter rotors},}\ }\href {\doibase 10.1098/rsif.2014.0585} {\bibfield
  {journal} {\bibinfo  {journal} {Journal of The Royal Society Interface}\
  }\textbf {\bibinfo {volume} {11}},\ \bibinfo {pages} {20140585} (\bibinfo
  {year} {2014})}\BibitemShut {NoStop}%
\bibitem [{\citenamefont {Cha}\ \emph {et~al.}(2020)\citenamefont {Cha},
  \citenamefont {Campbell}, \citenamefont {Popov}, \citenamefont {Stanczak},
  \citenamefont {Estep}, \citenamefont {Steager}, \citenamefont {Sung},
  \citenamefont {Yim},\ and\ \citenamefont {Bargatin}}]{Cha2020-1127}%
  \BibitemOpen
  \bibfield  {author} {\bibinfo {author} {\bibfnamefont {W.}~\bibnamefont
  {Cha}}, \bibinfo {author} {\bibfnamefont {M.~F.}\ \bibnamefont {Campbell}},
  \bibinfo {author} {\bibfnamefont {G.~A.}\ \bibnamefont {Popov}}, \bibinfo
  {author} {\bibfnamefont {C.~H.}\ \bibnamefont {Stanczak}}, \bibinfo {author}
  {\bibfnamefont {A.~K.}\ \bibnamefont {Estep}}, \bibinfo {author}
  {\bibfnamefont {E.~B.}\ \bibnamefont {Steager}}, \bibinfo {author}
  {\bibfnamefont {C.~R.}\ \bibnamefont {Sung}}, \bibinfo {author}
  {\bibfnamefont {M.~H.}\ \bibnamefont {Yim}}, \ and\ \bibinfo {author}
  {\bibfnamefont {I.}~\bibnamefont {Bargatin}},\ }\bibfield  {title} {\enquote
  {\bibinfo {title} {Microfabricated foldable wings for centimeter-scale
  microflyers},}\ }\href {\doibase 10.1109/JMEMS.2020.3013813} {\bibfield
  {journal} {\bibinfo  {journal} {Journal of Microelectromechanical Systems}\
  }\textbf {\bibinfo {volume} {29}},\ \bibinfo {pages} {1127--1129} (\bibinfo
  {year} {2020})}\BibitemShut {NoStop}%
\bibitem [{\citenamefont {{Van Treuren}}\ and\ \citenamefont
  {Wisniewski}(2019)}]{vanTreuren2019-121017}%
  \BibitemOpen
  \bibfield  {author} {\bibinfo {author} {\bibfnamefont {K.~W.}\ \bibnamefont
  {{Van Treuren}}}\ and\ \bibinfo {author} {\bibfnamefont {C.~F.}\ \bibnamefont
  {Wisniewski}},\ }\bibfield  {title} {\enquote {\bibinfo {title} {{Testing
  Propeller Tip Modifications to Reduce Acoustic Noise Generation on a
  Quadcopter Propeller}},}\ }\href {\doibase 10.1115/1.4044971} {\bibfield
  {journal} {\bibinfo  {journal} {Journal of Engineering for Gas Turbines and
  Power}\ }\textbf {\bibinfo {volume} {141}},\ \bibinfo {pages} {121017}
  (\bibinfo {year} {2019})}\BibitemShut {NoStop}%
\bibitem [{\citenamefont {{Van Treuren}}\ \emph {et~al.}(2020)\citenamefont
  {{Van Treuren}}, \citenamefont {Sanchez}, \citenamefont {Wisniewski},\ and\
  \citenamefont {Leitch}}]{vanTreuren2020-3955}%
  \BibitemOpen
  \bibfield  {author} {\bibinfo {author} {\bibfnamefont {K.~W.}\ \bibnamefont
  {{Van Treuren}}}, \bibinfo {author} {\bibfnamefont {R.}~\bibnamefont
  {Sanchez}}, \bibinfo {author} {\bibfnamefont {C.~F.}\ \bibnamefont
  {Wisniewski}}, \ and\ \bibinfo {author} {\bibfnamefont {P.}~\bibnamefont
  {Leitch}},\ }\bibfield  {title} {\enquote {\bibinfo {title} {Investigation of
  aeroacoustics and motor efficiency of a two-bladed stock and five-bladed
  propeller designs for static quadcopter applications},}\ }in\ \href {\doibase
  10.2514/6.2020-3955} {\emph {\bibinfo {booktitle} {AIAA Propulsion and Energy
  2020 Forum}}}\ (\bibinfo {address} {Virtual},\ \bibinfo {year} {Aug. 2020})\
  \bibinfo {note} {{AIAA} Paper 2020-3955}\BibitemShut {NoStop}%
\end{thebibliography}%

\end{document}